\documentclass{article}

\usepackage{PRIMEarxiv}

\usepackage[utf8]{inputenc} 
\usepackage[T1]{fontenc}    
\usepackage{hyperref}       
\usepackage{url}            
\usepackage{booktabs}       
\usepackage{amsfonts}       
\usepackage{nicefrac}       
\usepackage{microtype}      
\usepackage{lipsum}
\usepackage{fancyhdr}       
\usepackage{graphicx}       
\graphicspath{{media/}}     

\usepackage{amsmath,amssymb}
\usepackage{algpseudocode}
\usepackage[ruled, lined, longend, linesnumbered]{algorithm2e}
\usepackage[table]{xcolor}
\usepackage{booktabs}
\usepackage{multirow}

\usepackage{hyperref}

\usepackage{array}
\usepackage{tabularx}

\usepackage{listings}
\usepackage[inline]{enumitem}

\title{Management of Resource at the Network Edge for Federated Learning
}

\author{
  Silvana Trindade, Luiz F. Bittencourt, Nelson L. S. da Fonseca \\
  Institute of Computing\\
State University of Campinas \\
Campinas, Brazil\\
  \texttt{silvana@lrc.ic.unicamp.br, \{bit, nfonseca\}@ic.unicamp.br} \\
}

\begin{document}
\maketitle

\begin{abstract}
Federated learning has been explored as a promising solution for training at the edge, where end devices collaborate to train models without sharing data with other entities.
Since the execution of these learning models occurs at the edge, where resources are limited, new solutions must be developed.
In this paper, we describe the recent work on resource management at the edge, and explore the challenges and future directions to allow the execution of federated learning at the edge.
Some of the problems of this management, such as discovery of resources, deployment, load balancing, migration, and energy efficiency will be discussed in the paper.
\end{abstract}


\section{Introduction}\label{introduction}

Cloud computing provides on-demand delivery of resources and services over the Internet \cite{da2015cloud,DBLP:journals/csr/BittencourtGMFS18}.
Gartner has forecast that by 2025, 75\% of enterprise-generated data will be created and processed outside of centralized cloud data centers~\cite{gartner2018-edge}.
It has also forecast that more than 80\% of enterprise Internet-of-Thing (IoT) projects will include an Artificial Intelligence (AI) component by 2022, a significant increase when compared to that of 10\% in the last couple of years ~\cite{gartner2018}. 
Moreover, Cisco predicted that the number of mobile devices and connections will increase to 13.1 billion by 2023, with a compound annual growth rate of 8\% between 2018 and 2023~\cite{cisco2020}. 

However, the physical distance between users and the cloud servers introduces various issues, such as high latency, limited security and privacy, and  uncertainties in service reliability~\cite{8057196}. Executing tasks on cloud servers can jeopardize the support of strict latency requirements;
Issues related to the processing of AI and big data on the cloud have prevented the widespread adoption of AI in network applications and services~\cite{wang2020convergence, 9247530}.
 Edge computing has been proposed to overcome some of these challenges as it allows the deployment of machine learning, as well as mobile and real-time applications at the end-user~\cite{7912261,wang2020convergence}.
Indeed, edge computing allows the fulfillment of low latency requirements and real-time access to information~\cite{BITTENCOURT2018134}. 
Edge computing brings cloud services and resources to the edge, leading to several advantages, such as:
\begin{itemize}
    \item mobility support and location awareness, which  allows mobile users to access services from the closest edge server;
    \item proximity to the users, which decreases latency, and makes numerous mobile and real-time applications possible; 
    \item context-awareness, which supports the continuity of services in highly dynamic scenarios.
\end{itemize}

One significant advantage of AI models is the potential capacity for continuous  performance improvement as the number of examples (samples) increases.
However, the computational power required to execute such models can increase significantly as a function of the size of the dataset.
Executing learning models on the cloud implies the addition of long propagation delays in the response time of applications due to the need to send data from the user devices to the cloud servers. 
Moreover, such transfers can compromise the privacy of data and reduce security. 
Since devices have to share their data with the edge servers, this might discourage users from participating in the training of models, even when privacy mechanisms are applied.
To cope with these issues, recent studies have introduced processing based on machine learning at the edge, which helps achieve low-latency intelligent services~\cite{wang2018edge, wang2019adaptive} and~\cite{wang2020convergence}.
However, bringing learning models to the edge increases the complexity of managing a large distributed architecture.

To cope with this problem, the authors in~\cite{konevcny2016federated} and~\cite{mcmahan2017communication} have introduced the Federated Learning (FL) paradigm, which enables a large number of clients (edge devices and servers) to train local models with local data and then collaborate to the construction of a global model, shared by all the clients participating in the federation. 
Local data is stored locally, and collaborative training of a single machine learning model is carried out at the server without sharing local raw data~\cite{wang2019adaptive, konevcny2016federated}.

Edge computing can provide the infrastructure to execute FL.
Federated Edge Learning (FEEL)~\cite{tak2020federated} involves the ability to perform FL at the edge, where edge devices and edge serves can be either used to train models or produce the aggregation of local models.

FEEL brings a set of resource management challenges that impact the performance of FL tasks, since it defines how services and applications should be executed and how edge nodes should be  configured to receive the tasks.
These challenges include discovering  edge resources, deployment of FL tasks, traffic migration, load balancing, and energy consumption.
The constraints of limited bandwidth, uncertain device availability, and unbalanced data~\cite{kairouz2019advances} make resource allocation decisions crucial for an adequate execution of FL.

This paper focuses on the management of resources at the network edge to support federated learning.
Data imbalance, intense communication, task distribution, energy consumption, and high delay for migration are also addressed in this paper.
More specifically, the topics in edge resource management included are the following:
\begin{enumerate}
    \item Discovery: protocols for the discovery of edge resources to select resources for deployment.
    \item Deployment: solutions to deploy services and applications on federated edge networks.
    \item Load Balancing: mechanisms for balancing the workload at the edge.
    \item Migration: mechanisms for migrating tasks across edge nodes.
    \item Energy efficiency: solutions for increasing the energy efficiency without interfering with the performance of learning algorithms, including data training and inference.
\end{enumerate}

Both edge computing and FL have been the focus of great interest recently. 
However, resource management of edge devices to realize federated learning has not received much attention. 
This paper aims at filling this gap and suggests potential future investigations by pointing out open problems involving the mentioned aspects of edge resource management.

This paper comprises related work (Section~\ref{related-work}), discussions about edge computing (Section~\ref{edge-computing}), artificial intelligence (Section~\ref{ai-edge}), Federated Learning (Section~\ref{federated learning}), as well as discussions around various aspects of resource management, such as the discovery of edge resources (Section~\ref{discovery}), deployment of services and applications (Section~\ref{deployment}), migration (Section~\ref{migration}), resource allocation (Section~\ref{resource-allocation}), energy consumption (Section~\ref{energy}), and challenges (Section~\ref{challenges}).

\section{Related Work}\label{related-work}

This section presents a collection of papers on resource management at the edge and resource management for federated learning.
A more thorough presentation of work about specific topics will be provided in the body of the paper. 

\subsubsection*{Resource management for fog/edge computing}

A taxonomy of management for edge computing resources is presented in~\cite{tocze2018taxonomy}, where the authors divide resource management tasks into resource estimation, resource discovery, resource allocation, resource sharing, and resource optimization.
The relation between the Internet of Things (IoT) and  cloud/fog computing is surveyed in~\cite{DBLP:journals/iot/BittencourtISFM18}.
The authors discuss challenges of management and future directions, but they have not explored artificial intelligence as a solution for management problems AI  applications and services. 
In~\cite{hong2019resource}, the authors discuss the challenges of resource management for fog/edge computing. However, they have not discussed challenges related to the use of artificial intelligence algorithms at the edge.
In~\cite{wang2020convergence}, the authors survey different ways to explore machine learning algorithms at the edge but do not consider FL and only provide a list of advantages.
The authors explore the challenges associated with spectrum management for training and inference models in wireless communication. They recommend future investigation of communication bandwidth, noise, and interference since these are the main factors that influence  transmission quality.

 \begin{table}[!ht]
    \caption{Comparison of Related Work}
    \centering
    \fontsize{8}{10}\selectfont
    \renewcommand{\arraystretch}{1.3}
    \begin{tabular}{|p{1.2cm} | p{1.2cm} | p{6.8cm}| p{2.4cm}| p{2cm}|}
    \hline
 \textbf{Ref.} & \textbf{Year} & \textbf{Research Focus} & \textbf{Federated Learning} & \textbf{Management}\\
 \hline
 \cite{tocze2018taxonomy} & 2018 & Taxonomy of management for edge computing resources. & No & Yes \\\hline
 \cite{chen2018federated} & 2018 & Federated learning for vehicular networks & Yes & Yes \\\hline
 \cite{hong2019resource} & 2019 & Challenges of management of resources for edge/fog computing without consideration of artificial intelligence. & No & Yes\\\hline
  \cite{yang2019federated,wahab2021federated} & 2019, 2021 & Explored federated learning in a macro-perspective. & Yes & No \\\hline
 \cite{tran2019federated} & 2019 & Management challenges over wireless networks. & No & Yes\\\hline
 \cite{xu2020edge} & 2020 & Survey of edge intelligence. & &\\\hline
 \cite{aledhari2020federated} & 2020 & Federated learning for applications and use cases. & Yes & No \\\hline
 \cite{wang2020convergence} & 2020 & Edge intelligence and intelligent edge explored in a macro-perspective.  & No & No \\\hline
 \cite{9220170} & 2020 & Challenges and issues of device selection, resource allocation, and updates aggregation in federated edge learning.  & Yes & No \\\hline
 \cite{lim2020federated} & 2020 & Federated edge learning. & Yes & No\\\hline
 \cite{9247530} & 2020 & Exploration of the interaction of edge devices with a central server. & & \\\hline
 \cite{yang2021federated} & 2021 & Federated learning in 6G. & Yes & No \\\hline
 Our & 2021 & Resource management for federated learning at the edge.   & Yes & Yes\\\hline
\end{tabular}
\label{tab:surveys-related}
\end{table}
The authors in~\cite{xu2020edge} surveyed edge intelligence, and open challenges in edge offloading, edge caching, edge training, and edge inference.
Resource management was briefly discussed throughout the paper, but FL was not considered.

\subsubsection*{Resource management for federated learning}

The architecture and applications of FL are the focus of the discussions in~\cite{yang2019federated, wahab2021federated}.
In~\cite{yang2019federated}, the authors divide FL into vertical, horizontal, and transfer learning.
Different from~\cite{yang2019federated}, the authors in~\cite{wahab2021federated} explore different aspects of federated learning, including basic concepts of federated learning, a comparison between federated learning and other distributed learning approaches, and emerging communication technologies (including edge computing).
However,  challenges in resource management are not explored.
In~\cite{chen2018federated}, the authors survey existing mechanisms and the challenges of federated learning for vehicular networks, including a few mentions of resource management.
The use of federated learning in smart cities is surveyed in~\cite{jiang2020federated}, where challenges and opportunities are discussed, including deployment and energy consumption.

The management of federated learning over wireless networks is explored in~\cite{tran2019federated}.
The solution proposed by the authors includes a model for implementing FL and a discussion of how it affects the energy consumption of user devices, computational time, and communication latency in the execution of FL applications.

The interaction between the global server and edge devices when employing FL is introduced in~\cite{9247530}.
The authors use a game strategy to motivate users to participate in local training.
The authors also list some of the challenges faced when executing FL at the edge, such as the security of the global model and the optimization of computational and communication resources.

In~\cite{aledhari2020federated}, the authors discussed FL for specific applications and use-cases, and pointed out some of the challenges related to data imbalance, missing features and values, and security. 
However, they have not explored challenges related to network operation.   
Design issues and some of the challenges faced in federated edge learning are introduced in~\cite{tak2020federated}, with the authors describing issues related to model aggregation
node selection, energy consumption, and resource allocation.
The authors in~\cite{lim2020federated} and \cite{yang2021federated} focus on 
FL in mobile edge computing.
In~\cite{lim2020federated}, existing solutions for the implementation of FL at the edge are reviewed, while in~\cite{yang2021federated}, the challenges of executing FL in the sixth cellular generation (6G) are discussed, emphasizing open problems in wireless communication that can be addressed by the use of FL.
 
In~\cite{9220170}, the authors have explored design issues and challenges of federated edge learning. They proposed a generic framework for adding data properties on the edge.
Their solution divides the algorithm designed for federated edge learning into device selection, resource allocation, and update aggregation. They suggest that, in order for FL to be executed at the edge, both resource and data need optimization.

 \begin{figure}[!ht]
    \centering
    \includegraphics[scale=0.16]{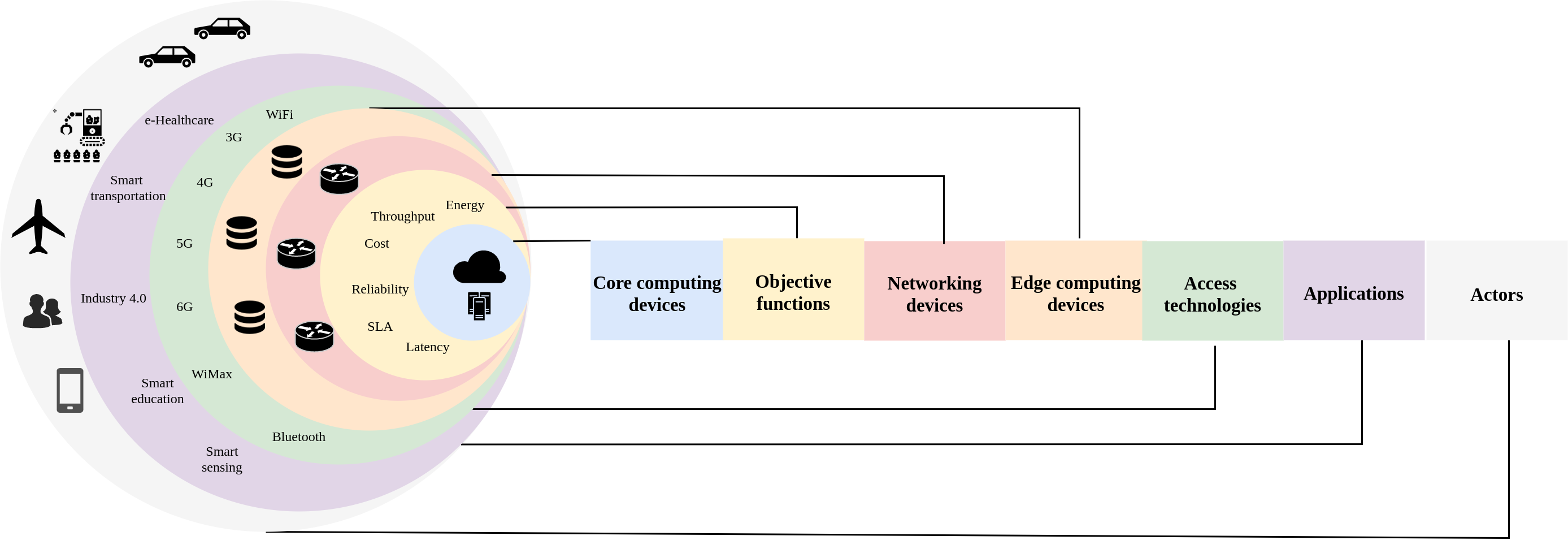}
    \caption{Layered architecture of the edge computing paradigm~\cite{8291113}.}
    \label{fig:edge-layer}
\end{figure}
In contrast to what is found in the literature, in this paper, we focus on defining the resource management problem for federated learning and discuss challenges that arise when managing edge resources running federated learning tasks.
Our survey focuses on the management of resources at the edge in scenarios where federated learning is employed.
Our contributions can be summarized as follows:
\begin{itemize}
     \item We survey the recent work on edge computing combined with artificial intelligence, specifically, supervised machine learning techniques. We discuss the challenges that are faced on edge computing when executing machine learning at the edge.
     \item We survey concepts of federated learning and how it works at the network edge, focusing on the training process for this technique. Recent work and challenges are discussed throughout the paper. 
     \item We survey recent researches on resource management. We have divided it into five components: discovery, deployment, load balancing, migration, and energy consumption. We discuss and describe each of these and how they interact with federated learning.
     \item We also discuss challenges and issues of resource management, which must be dealt with to provide a suitable environment for federated learning at the edge. Directions for future work are also presented in this survey.
 \end{itemize}
 
\section{Edge Computing}\label{edge-computing}

Edge computing is a distributed computing paradigm that brings computation and data storage capabilities to the edge of the network, closer to the location where it is needed, thus reducing delays and bandwidth utilization in the network core.
This paradigm reduces computation and data offloading to the cloud.

There are two types of edge nodes n edge computing: edge servers and edge devices with different computational capacities.
An edge server can be a Personal Computer (PC) or a half-rack built for processing Information Technology (IT) workloads (micro data center).
Such edge servers are usually located near or on the routers or at access points.
Edge devices are typically mobile phones, vehicles, Internet-of-Things (IoT) devices, and robots with limited computational resources.
These devices usually access edge servers to execute different tasks, creating a distributed environment.
Edge computing may consist of the following~\cite{wang2020convergence}:
\begin{itemize}
    \item \textit{Cloudlet and Micro Data Centers} - a network architecture element that combines mobile computing and cloud computing. It represents the middle layer of a three-tier architecture (mobile devices, micro cloud computing, and the cloud). 
    \item \textit{Fog computing} - a fully distributed multi-tier architecture for cloud computing with billions of devices and large-scale cloud data centers. The deployment of a fog is targeted at specific geographic areas; it is designed for applications that require real-time responses with strict latency requirements, such as highly interactive and Internet of Things (IoT) applications.
    \item \textit{Mobile Edge Computing/Multi-Access Edge Computing (MEC)} -  furnishes computing capabilities and service at the edge of cellular networks. It is designed to provide low latency, context and location awareness, and large bandwidth.  
\end{itemize}

Figure~\ref{fig:edge-layer} depicts the elements of an edge computing architecture: core computing devices, networking devices, edge computing devices, access technology, applications, and actors. 
Smart devices, vehicles, and individuals play the role of actors who try to access different on-demand applications, such as smart sensing, smart education, e-healthcare, and smart transportation. 
To access these services, the actors need to use the available access technologies to connect to the computing platform ~\cite{8291113}.
The access technology relays the user application/service requests to the nearby nano-Data Centers (nDCs) and micro-data centers (mDCs).
Edge servers have a limited computational capability; hence, they forward the computing-intensive requests to the core computing elements. 
Such requests are routed by networking devices, such as routers and switches, for further processing.
The entire computing architecture tends to address various requirements in the Service Level Agreement (SLA), such energy consumption, response time, latency, availability, and throughput.

In edge computing, tasks can be executed either in isolation or in a collaborative or federated manner, as described next.

\subsection*{Collaborative Edges}

Artificial Intelligence algorithms, such as Deep Learning (DL), involve intensive computation tasks that most edge devices cannot handle~\cite{wang2020convergence}. Moreover, a single device may not be able to store enough data for proper learning.
Edge computing can, however, potentially cope with these problems by offloading the computation of learning algorithms from the edge devices to the edge servers. 
In this way, edge servers can become an extension of the cloud infrastructure to deal with massive learning tasks, and perform location-based learning.
If an edge server does not have enough data, it needs to offload the model inference/training to cloud data centers or download the necessary data from the cloud. 
In either case, data processing latency increases. 

To support intelligence at the edge, tasks can be divided into sub-tasks and distributed across various edge nodes for collaboration.
Such  collaboration among edge nodes can happen either vertically or horizontally.
In vertical collaboration, edge nodes and cloud servers share a task, whereas in horizontal collaboration, the whole task is divided and executed at the edge, with only the results sent to cloud servers for long-term storage or further processing.
These two different approaches involve distinct latency and communication with the cloud.

Vertical collaboration brings certain advantages.
In vertical collaborative edge computing, the edge performs data pre-processing and preliminary learning upon offloading of learning tasks, and this intermediate data is then sent to the cloud for further computation. 
The hierarchical structure of the machine learning algorithm allows vertical collaboration. 
This is the case when all layers of neural networks are profiled on the edge device, and the edge server makes predictions on the basis of massive data~\cite{wang2020convergence}. 

In horizontal collaboration, edge devices can collaborate without involving  cloud servers to process  resource-hungry machine learning applications. 
The trained machine learning models or  the whole task can be partitioned and allocated to multiple end devices or edge servers to accelerate computation.

\subsubsection*{Federated Edges}

As of today, most edge resources are configured in an ad-hoc manner and may or not be publicly available, depending on the use case. 
They may also be sparsely distributed.
Such ad-hoc, private, and sporadic edge deployments are less useful in transforming the global Internet. 
The benefits of edge computing should be equally accessible to ensure computational fairness, as well as to connect billions of devices on the Internet.
However, very little effort has been made to employ edge computing on a global scale.
Federating edge devices across multiple geographic regions can create a global edge-based fabric that decentralizes data center computation.

Recently, federated edge deployment has achieved federation by employing a load-balancer at each edge-aware of the end-to-end latency between edge systems~\cite{8855656}.   
In~\cite{8855656}, the authors use a federated resource allocation scheme to achieve intelligence in information processing for oil industry plants. 
The authors in~\cite{7561038} propose a federated architecture combined with a distributed Edge Cloud-IoT platform, supporting a cross-layer approach for IoT service provisioning and management over distributed mobile edge cloud environments. 
The approach was used for scenarios with only two smart health applications/services  and no real-world scenarios, including a large number of edge nodes was conducted. 

In~\cite{afolabi2019dynamic}, the authors proposed a deployment framework for federated mobile networks.
They employed a dynamic auto-scaling approach to efficiently scale the network resources, making the orchestration of system resources more efficient~\cite{afolabi2019dynamic}.

\begin{figure}[!ht]
    \centering
    \includegraphics[scale=0.34]{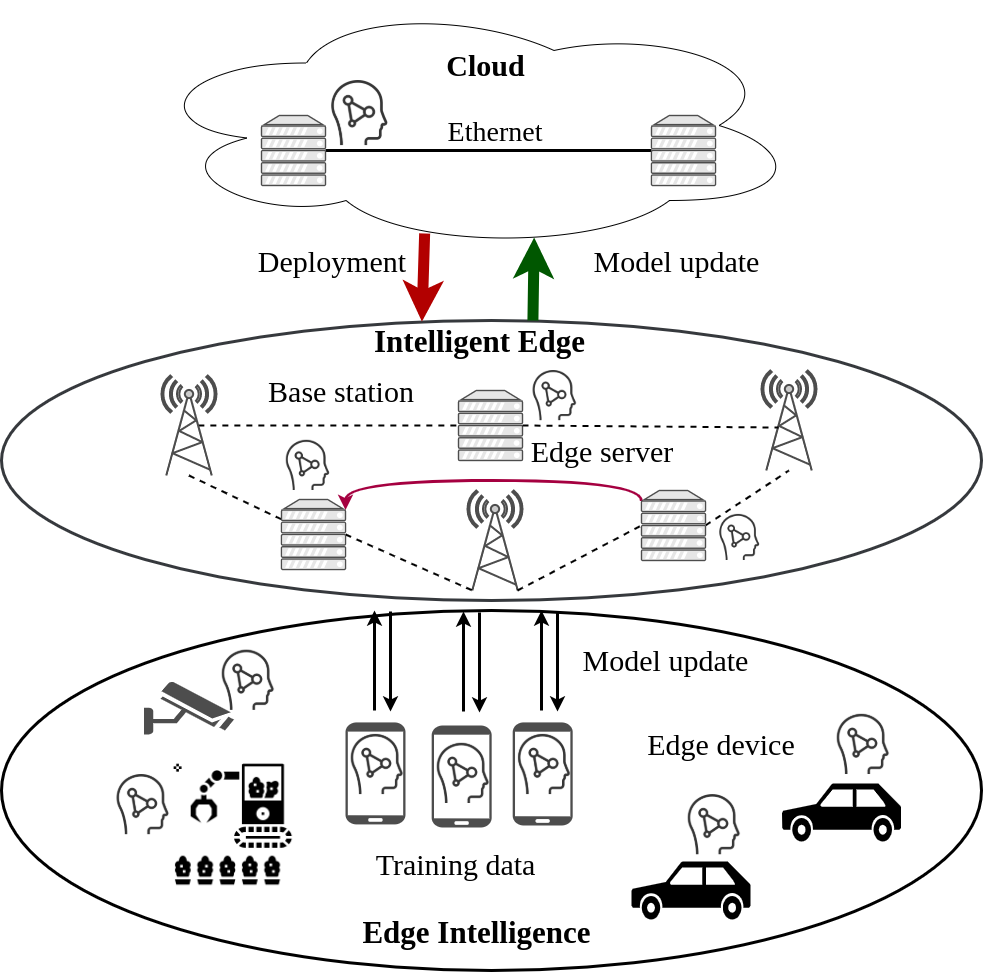}
    \caption{Edge intelligence and intelligent edges~\cite{wang2020convergence}.}
    \label{fig:ei_ie}
\end{figure}

Federating edge resources also introduces  several  networking challenges related to the scalability of user demands are constantly changing due to their mobility and the dynamics of the environment. 
Moreover, the Quality-of-Service (QoS) of the deployed applications and services need to be assured.
For example, in industry 4.0, the data sent to different edge nodes used for transportation may require adaptation to changes in traffic conditions. 
One possible way to deal with a dynamic environment is the integration of edge computing and Software-Defined Network (SDN)~\cite{prados2020learnet}.

\section{Artificial Intelligence at the Edge}\label{ai-edge}

Artificial Intelligence (AI) builds intelligent machines capable of carrying out tasks the way humans do. 
Apple Siri and Google AlphaGo are just examples of such powerful technologies. 
 AI systems typically exhibit at least some of the following characteristics associated with human intelligence: planning, learning, reasoning, problem-solving, knowledge representation, perception, motion, manipulation, and to a lesser extent, social intelligence and creativity. 
After 2010, AI has experienced a big surge due to the advances in deep learning, a method that has achieved human-level accuracy in some areas~\cite{zhou2019edge}.
Edge computing aims at coordinating a multitude of collaborative edge devices and servers to process the data generated in the proximity of the user.
Edge computing and AI can synergize to create an enhanced computational environment to allow new services and applications~\cite{zhou2019edge}.

Edge computing has gradually been combined with AI, each benefiting the other in terms of intelligent edge and the realization of edge intelligence (Fig.~\ref{fig:ei_ie}).
Edge intelligence is defined as the incorporation of artificial intelligence for user applications and services, such as e-health applications and recommendation systems to be executed at the edge.
An intelligent edge refers to an edge that uses artificial intelligence techniques to make decisions impacting on resource usage at the edge, such as caching, offloading, scheduling, resource allocations, or load balancing; all of  these technologies are crucial to the deployment of learning tasks at the edge.
Edge intelligence and intelligent edges are not fully independent; edge intelligence is the goal. Machine learning services on intelligent edge serves as the key for edge intelligence. 
Intelligent edges can support high service throughput and great resource utilization for edge intelligence~\cite{wang2020convergence}.

\begin{figure}[!ht]
    \centering
    \includegraphics[scale=0.22]{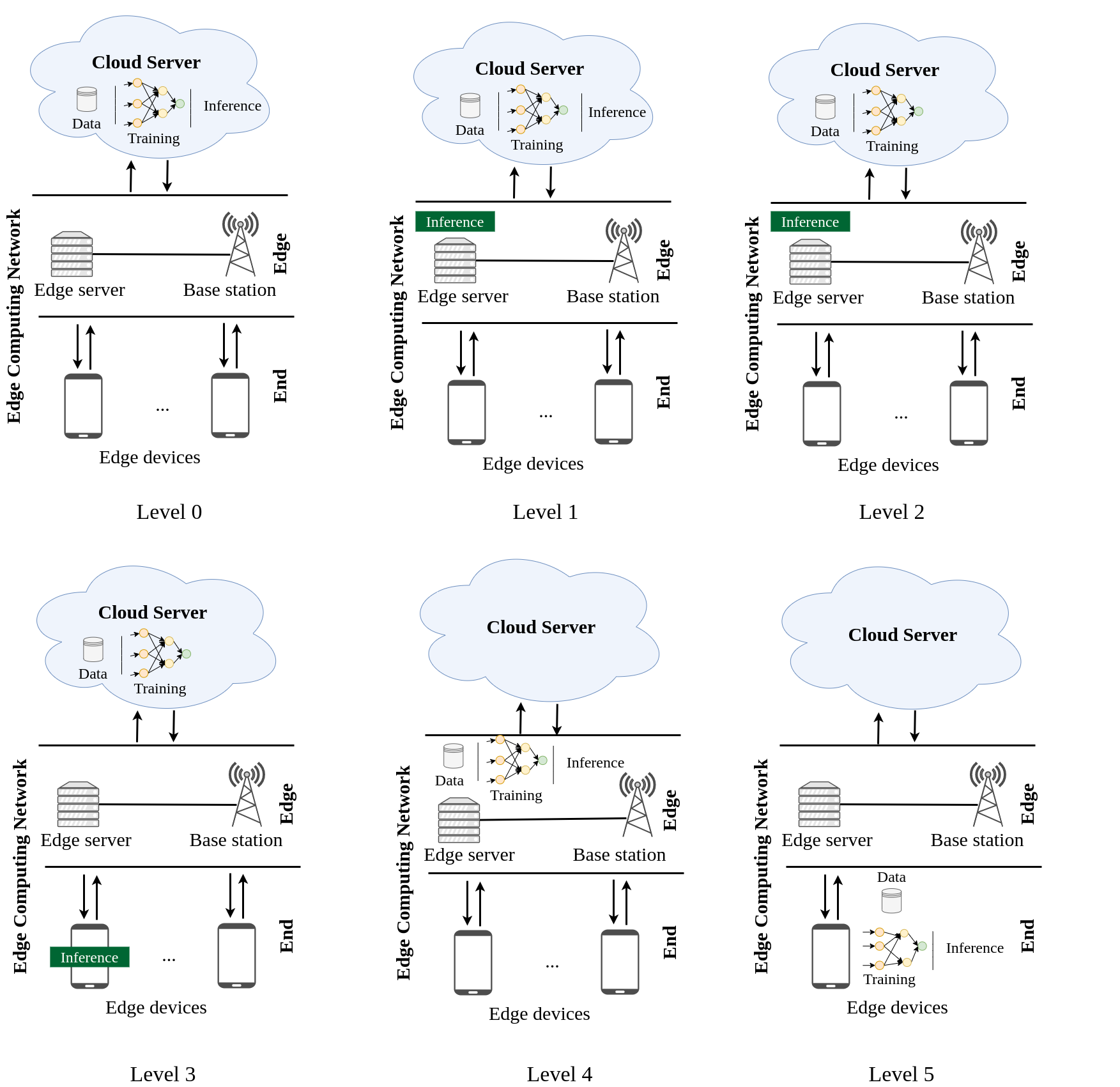}
    \caption{Processing in classified levels of edge intelligence on the basis of the location of resource used~\cite{zhou2019edge}.}
    \label{fig:edge-cloud-ia-levels}
\end{figure}

Edge intelligence exploits available data and resources across the hierarchy of end devices, edge servers, and cloud datacenters to optimize the overall performance of training and inference of a learning model.
In~\cite{zhou2019edge}, edge intelligence is classified into different levels according to the interaction between cloud, edge, and user devices (Figure~\ref{fig:edge-cloud-ia-levels}) described as follows:
\begin{enumerate}
    \item \textit{Level 0 - Cloud Intelligence}: Training and inference carried out fully on the cloud. 
    \item \textit{Level 1 - Cloud–Edge Co-inference and Cloud Training}: Training model carried out on the cloud, and inference is performed by edge-cloud cooperation with data partially offloaded to the cloud.
    \item \textit{Level 2 - In-Edge Co-inference and Cloud Training}: Training of the model is carried out on the cloud, while model inference is carried out at the edge, with data fully or partially offloaded to the edge nodes or nearby devices (e.g., via device-to-device communication). 
    \item \textit{Level 3 - On-Device Inference and Cloud Training}: Training models are carried out on the cloud, while inference takes place on a local edge device so that no data is  executed at the edge.
    \item \textit{Level 4 - All In-Edge}: Both training and inference are carried out using the in-edge approach. 
    \item \textit{Level 5 - All On-Device}: Both training and inference are carried out using an on-device approach.
\end{enumerate}

There are many different ways in which intelligence can be incorporated on the edge, and as a consequence, powerful learning techniques are required in large scale systems with thousands of devices being trained and inferences being made.
One of these promising techniques is Federated Learning (FL)~\cite{yang2019federated}, where devices at the edge of the network are used to support the training of the learning model without sending raw data to a central server.
\begin{figure}[!ht]
    \centering
    \includegraphics[width=0.65\textwidth]{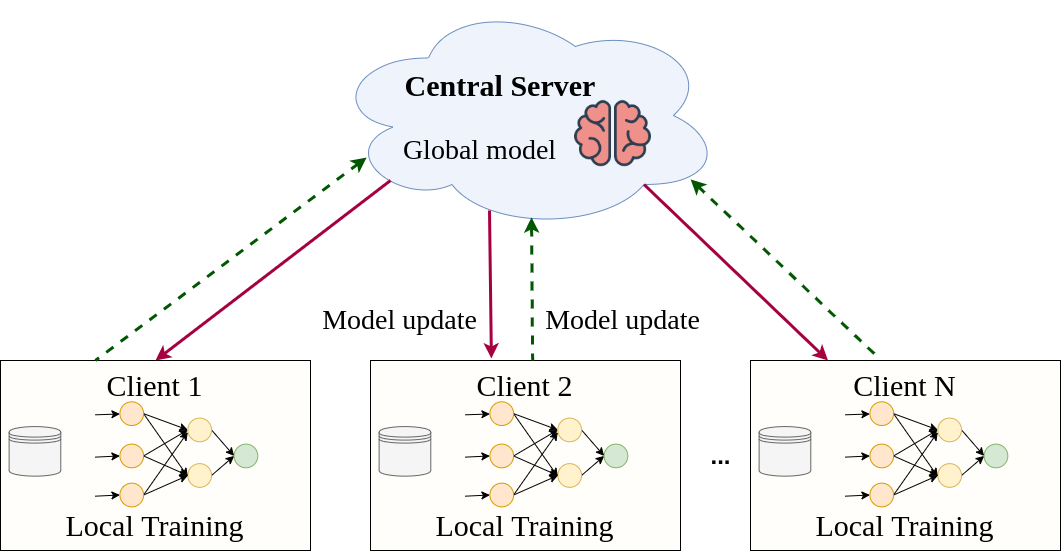}
    \caption{Centralized model in federated learning.}
    \label{fig:c-fl}
\end{figure}
FL can be implemented in three different ways (Figure~\ref{fig:edge-cloud-ia-levels}): edge devices being responsible for training with cloud servers aggregating the trained model parameters; edge devices being responsible for training and edge servers for the aggregation parameters; and edge devices being responsible for both training and aggregation so that no interaction with cloud servers is necessary during model training.

As more intelligence is placed at the edge, the amount and path length for data offloading is reduced; as a result, transmission latency  decreases, data privacy increases, and the usage of Wide Area Network (WAN) bandwidth is reduced. 
However, this comes at the cost of increased complexity in the management of resources and energy consumption~\cite{zhou2019edge}.
There is no optimal approach for processing all  machine learning applications. 
It is, therefore, necessary to analyze multiple criteria such as latency, privacy, energy consumption, and bandwidth to determine the most advantageous approach for each application. 

\section{Federated Learning}\label{federated learning}

Edge devices such as mobile phones and sensors produce a wealth of data at the edge.
However, due to data privacy, it is not always possible to transmit all of these data from the edge devices to a cloud data center to build a machine learning training model.
Federated Learning (FL)~\cite{konevcny2016federated,mcmahan2017communication} has been proposed to address these constraints.
Federated learning is a distributed machine learning (ML) (Figure~\ref{fig:c-fl}) that preserves privacy while training learning models based on data originated from multiple clients.
Rather than sending the locally generated raw data to a centralized data center for training, FL leaves the raw data on the clients (e.g., mobile devices) and trains a shared model, on both edge and cloud servers, by aggregating the model trained locally by the clients on the cloud.

Federated learning is a distributed way for executing 
machine learning models that are characterized by the following features~\cite{wang2019adaptive}:
\begin{itemize}
    \item there are two or more parties interested in jointly building an ML model. Each party holds its  own raw data.
    \item In the model training process, the data held by a party does not leave the device.
    \item The model can be partially transferred from one party to another under an encryption scheme, so that  data cannot be re-engineered.
    \item The performance of the resulting model is a good approximation of the ideal model built with all the data of all the parties in the federation.
\end{itemize}

The federated learning problem involves learning a single, global statistical model from data stored on tens to potentially millions of remote devices~\cite{yang2019federated}.  
The goal of  FL is to learn a model generated by the data from edge devices that are stored and processed locally.

\subsection*{Applications}

Federated Learning (FL) allows promising applications in many different fields such as sales, finance, healthcare, education, urban computing, edge computing, and blockchain~\cite{brisimi2018federated, xu2020federated, yang2018applied, samarakoon2019distributed, duan2019astraea, kang2020reliable, rieke2020future, xu2021federated}.
In these many applications, it is possible to aggregate all the data directly on the training Machine Learning (ML) models for a variety of reasons, such as communication delays, overheads, network limitations, and data privacy. 
\begin{figure}[!ht]
    \centering
    \includegraphics[width=0.99\textwidth]{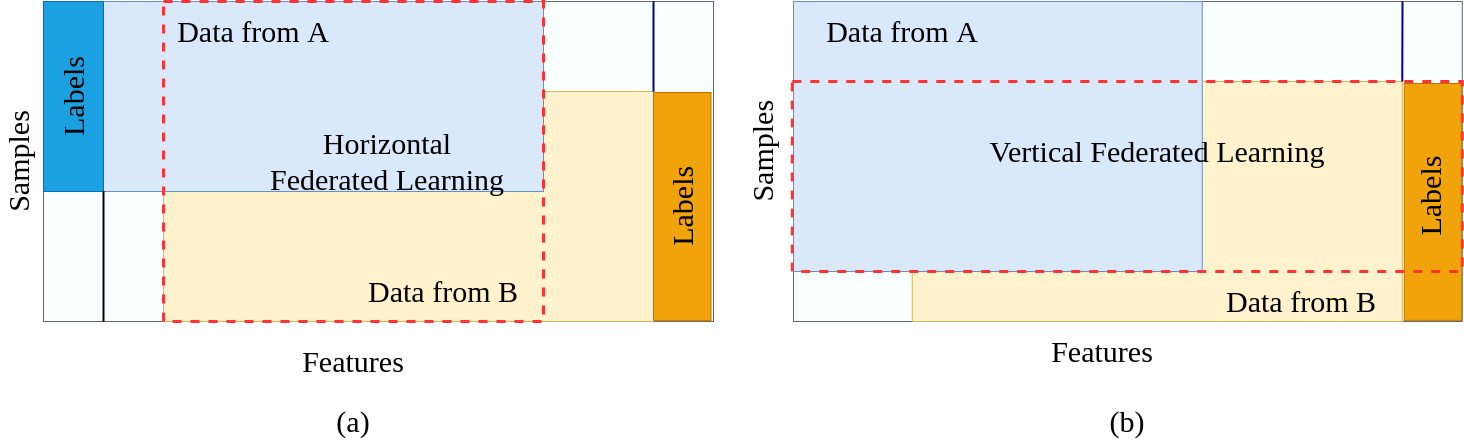}
    \caption{Data partition: horizontal federated learning versus vertical federated learning.}
    \label{fig:hfl-vfl}
\end{figure}

In the financial industry~\cite{yang2019ffd}, for example, government regulations are designed to protect investors against mismanagement and fraud, as well as maintain the stability of the financial sector, preserving the privacy and security of the user. 
Federated learning can  deal with user privacy, data security, and data storage while providing technical support for  building  a cross-enterprise, cross-data and cross-domain ecosystem for big data and AI.

In healthcare~\cite{xu2020federated, brisimi2018federated}, AI has been adopted to reduce labor costs and human errors in applications as cardiology and radiology.
These applications were designed to help diagnose diseases (e.g., heart diseases and cancer cells), potentially improving the diagnosis provided by doctors.
However, such applications must also deal with the privacy of data as well as misleading diagnoses.
By employing FL, medical institutions can share their data while still complying with privacy protection regulations.

Mobile Edge Computing (MEC)~\cite{samarakoon2019distributed, duan2019astraea} makes it possible to process data instead of sending it to cloud servers. This promotes the computation of large models in a federated way while coping with strict latency requirements.
Since FL can be a solution for network problems (intelligent edge), the authors in~\cite{wang2020federated} proposed  federated learning to improve the orchestration of computation at the edge.

\subsection*{Data Partitioning}

There are three ways to process data using FL: Horizontal Federated Learning (HFL), Vertical Federated Learning (VFL), and Federated Transfer Learning (FTL)~\cite{hard2018federated, lim2020federated, liu2020secure}.
HFL (Figure~\ref{fig:hfl-vfl} (a)) applies to scenarios where participant datasets share the same feature space but differ in sample spaces.
HFL requires all participating entities to share the same feature space, while VFL requires entities to share the same sample space~\cite{lim2020federated, feng2020multi}. 
For example, two regional banks may have similar business models but very different clients, so the intersection of the sets of their users is small.
Such a situation is typical of a Business-to-Costumer (B2C) relationship.

For Business-to-Business (B2B), VFL~\cite{liu2020asymmetrically, feng2020multi} (Figure~\ref{fig:hfl-vfl} (b)) is more suitable since it leverages the heterogeneous feature spaces of distributed datasets maintained by organizations to improve machine learning models without exchanging and exposing the private data of their clients.
One example of B2B would be a hospital collaborating with pharmaceutical companies, making use of the medical records of patients in treating chronic diseases to reduce the risk of future hospitalization.
In these cases, the two entities have different goals and datasets, but the dataset is composed of data from the same patients~\cite{lim2020federated}.

In real-world scenarios, it is common that not enough features or samples exist for sharing with federated clients; an FL model can still be created using transfer learning, in which the knowledge is transferred across the participating entities, in order to improve the performance of a shared model.
FTL clients share only a partial overlap in the user or feature space but leverage existing transfer learning techniques to build learning models collaboratively~\cite{yang2019federated}.
FTL can be helpful in heterogeneous scenarios where datasets may share only a handful of samples and features, even though the size of the datasets varies significantly. Not all data have labels (or there are limited labels)~\cite{zhao2018federated}.

\subsection*{Model Update Aggregation}

In~\cite{zhao2018federated}, the authors introduced the \textit{FederatedAveraging} ($FedAvg$) algorithm~\cite{mcmahan2017communication,chen2016revisiting} to combine  Stochastic Gradient Descent (SGD) parameters of local models in a server that performs the model aggregation.
In the $FedAvg$,~"each party uploads a clear-text gradient to a coordinator (or trusted dealer, or a parameter server) independently, then the coordinator sends an updated clear-text back to each party"~\cite{zhao2018federated}.
This is possible since, in gradient descent methods, the objective function can be decomposed into differentiable and linearly separable functions.

In~\cite{chen2019asynchronous} and~\cite{xie2019asynchronous}, the authors introduced the $FedAsync$ algorithm that uses asynchronous aggregation.
Newly received local updates are weighted according to their staleness. Stale updates received from stragglers are weighted less based on how many rounds elapsed.
The $FedAsync$ is more suitable for heterogeneous scenarios, but there are still open  communication, efficiency, and security issues that prevent its broad adoption in the real-world.

\subsection*{Federated Learning at the Edge}

In Mobile Edge Computing (MEC), some issues that need to be overcome to deploy federated learning applications~\cite{8770530}: 
\begin {enumerate}
\item the training data is massive when considering massive edge devices,  increasing the burden of uplink wireless channels; 
\item the training data, which should be uploaded to edge nodes or cloud, is privacy-sensitive; 
\item if the training data is transformed for privacy consideration, server-side proxy data is less relevant than on-device data. 
\end{enumerate}
The authors also listed two deficiencies when they execute Deep Reinforcement Learning (DRL): the computation capability of an edge device is relatively small, and it can take a long time to train the DRL agent using massive data. Moreover, the training process of a DRL agent may consume the additional energy of edge devices.
A collaborative edge framework that executes DRL using the FL processing is proposed to address these issues.

In~\cite{8770530}, the authors studied the DRL using the FL paradigm. 
Edge devices perform training with their  models and then send the resulting model to be aggregated with other models to compose a global model. 
However, their proposal does not explore specific methods for optimizing learning computation tasks at the edge.
Their work also does not include  optimizing  edge nodes to collaborate and how to allocate resources to intelligence tasks considering deadlines, scales, CPU, and memory requirements.

Another way to classify federations in federated learning considers the number of participating devices~\cite{kairouz2019advances}.
Cross-silo federated learning  typically has up to 100 edge devices, while the cross-device can have thousands of devices.
In cross-silo, the bottleneck can be either the computation or communication, whereas in cross-device communication is typically the bottleneck  due to the use of Wi-Fi or other types of connection with smaller capacity. 
Another difference between these two is the availability of clients: while in cross-silo clients are available most of the time, each client in cross-device can be intermittently available.
However, cross-silo can either use vertical or horizontal partitions while cross-device uses horizontal partition only.

In~\cite{wahab2021federated}, the authors discuss a collection of advantages resulting from using federated learning and edge computing together, including 
\begin{enumerate*}[label={\alph*)}]
    \item the reduction of latency and bandwidth usage since client's data are not transmitted to the server;
    \item security is increased since the information change decreases;
    \item scalability increases since one model is trained in  collaboratively;
    \item reliability increases because of the reduction of network problems and 
    \item transfer learning empowerment.
\end{enumerate*}

However, placing model training on the edge introduces new challenges and calls for edge devices with more powerful processors to train complex local models, as well as proper management of learning tasks.
Moreover, AI technologies and edge computing will not be developed in isolation but will move toward integrated development.

Several challenges need to be addressed to realize FL~\cite{kobayashi2019radio}:
\begin{figure}[!ht]
    \centering
    \includegraphics[width=0.95\textwidth]{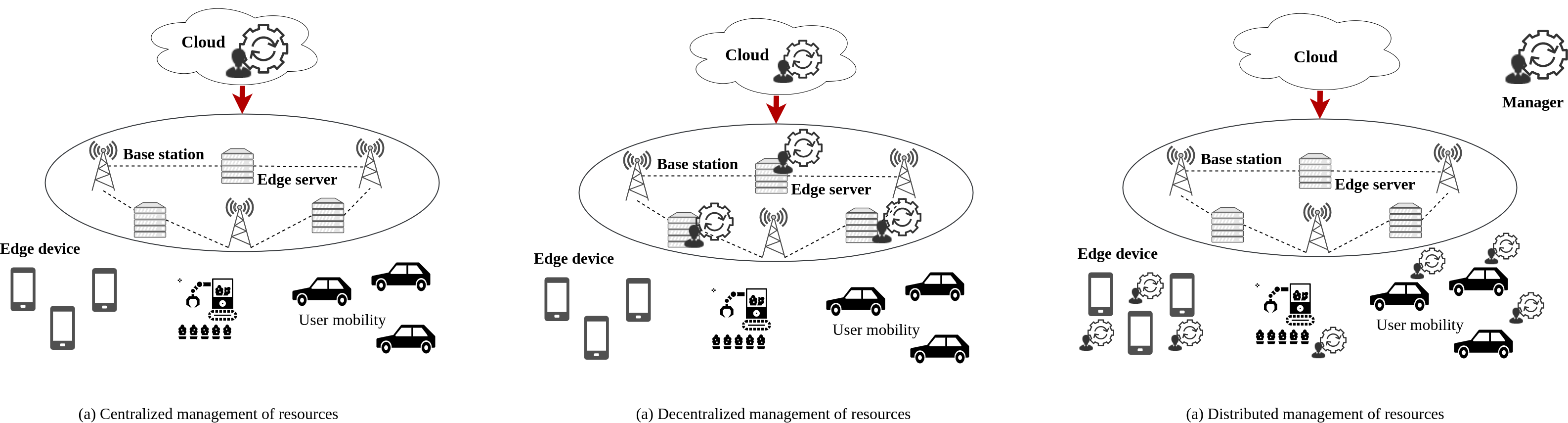}
    \caption{Types of management of resources in edge computing: centralized, decentralized, and distributed.}
    \label{fig:manager_types}
\end{figure}
\begin{itemize}
    \item \textit{Privacy}: In FL, raw data never leaves the users' device. Nonetheless, having more users involved in one collaborative model increases the risk of launching inference attacks that aim at inferring sensitive information from users’ training data.
    
    \item  \textit{Communication}: FL reduces the amount of information that needs to be transmitted over the network. However, since machine learning models are trained collaboratively, several model updates need to be communicated between the clients and the server in several iterations, which poses trade-offs in accuracy and communication costs.
    
    \item \textit{Latency}: FL should be appropriately managed to maintain low latency and low waiting times.
    
    \item \textit{Statistical Heterogeneity}: given that the training data on each client device depends on its  usage patterns, the local dataset of one client in FL is not expected to be representative of the overall data distribution. Similarly, as clients use their services or applications in varying degrees, the local datasets across clients tend to have different sizes.
    
    \item \textit{Massive Distribution}: it is possible that the number of clients  participating in the federated training be significantly larger than the average number of training samples per client.
    
    \item \textit{Connectivity}: in FL, client devices are frequently offline or have limited connectivity in FL. The selection of clients to participate in the federated training might consider the connection quality conditions (e.g., local time zone and the device being charged)
\end{itemize}

 Since  clients  send updated model parameters to the server in a typical round with large models, this step will  likely to become a bottleneck for computation, especially when network connections are unreliable. 
 The main challenges that affect the execution of FL at the edge involve the optimization of resource usage and communication, with a trade-off between frequent model updates and network utilization.
The communication issue is a consequence of the need for processing until the learning model converges.

\section{Management of Resources}\label{management}

Figure~\ref{fig:manager_types} illustrates three types of management of resources in edge computing: centralized, decentralized, and distributed.
In a centralized approach, information is collected from all edge nodes and sent to a central node (e.g., cloud server), which will then process the data and make decisions.
This approach is recommended when a single entity can reliably collect information about the whole system.
In a decentralized approach,  a central node and other nodes (e.g., edge nodes) can make decisions.
In a distributed approach, decisions can be taken independently by any of these entities.

The management of resources calls for monitoring the availability of resources in different layers of the network, such as the availability of incoming user devices.
A solution for this monitoring is the addition of a layer of edge nodes between the cloud and the user devices. However, it introduces significant monitoring overhead.
This becomes even more challenging when clusters of edge nodes from different geographical locations are to be federated to create a global architecture. 

The following sub-sections will discuss the most relevant issues in resource management for federated learning, such as the discovery of edge nodes, deployment of services and applications, load balancing, energy efficiency, and migration. 
Each section includes an introduction to the topic and a description of related work that  has already been done.

Table~\ref{tab:m-frameworks} shows the existing frameworks and platforms for the management of resources at the network edge.
Currently, most existing edge frameworks~\cite{wang2017enorm, varghese2017edge, amento2016focusstack, 9265252, zeng2019resource} do not have mechanisms that allow FL to run effectively at the edge or they focus on a single research area~\cite{clara2019,fate2019}. 
The Federated AI Technology Enabler (FATE)~\cite{fate2019} is an open-source project intended to provide a secure computing framework to support a federated AI ecosystem focused on industrial solutions, while NVIDIA Clara focuses on healthcare.
The frameworks introduced in~\cite{wang2017enorm, varghese2017edge} and~\cite{amento2016focusstack} also have limitations, such as lack of multi-tenancy, lack of support for FL, and lack of support for heterogeneous scenarios.

A resource management framework for 5G networks was explored in~\cite{boudi2021ai}.
The authors proposed a resource manager based on artificial intelligence and divided the tasks between edge and cloud.
Their solution alleviates the scheduling and placement problem in large-scale systems~\cite{boudi2021ai}.
The authors used real testbeds for  validating of the framework.

\begin{center}
    \begin{tabularx}{0.95\textwidth} { 
      | >{\raggedright\arraybackslash}X 
      | >{\raggedright\arraybackslash}X
      | >{\raggedright\arraybackslash}X 
      | >{\raggedright\arraybackslash}X |}
     \hline
     \multicolumn{4}{|c|}{\uppercase{Frameworks and platforms for resource management}}\\\hline
     \textbf{References} & \textbf{Domain} & \textbf{Solution} & \textbf{Machine Learning} \\\hline
     \cite{wang2017enorm} & Edge computing & Discovery and deployment & No\\\hline
     \cite{kubeedge} & Edge computing &  Discovery & No\\\hline
    \cite{fate2019} & Federated learning &  Deployment & Yes\\\hline
     \cite{liu2016paradrop} & Edge computing &  Deployment  & No \\\hline
    \cite{clara2019} & Machine learning & Deployment & Yes\\\hline
    \cite{zeng2019resource} & Intelligent edge & Orchestration of resources and migration & No \\\hline
     \cite{amento2016focusstack} & Edge computing & Discovery and deployment & No\\\hline
     \cite{WANG2021102016} & Edge intelligence &  Deployment & Yes \\
    \hline
    \end{tabularx}
    \label{tab:m-frameworks}
 \end{center}
\subsection{Edge Nodes Discovery}\label{discovery}

The discovery of edge nodes (routers, base stations, switches, and dedicated low power computational devices) involves the identification of the geographical location of edge nodes and their status so that cloud/edge servers can deploy services and applications on them~\cite{varghese2017edge}. 

The problem of resource discovery takes place on either individual or collective levels. 
On the individual level, potential edge nodes that can provide computing resources must be visible  to applications running on user devices and to cloud servers.
On the collective level, a collection of edge nodes in a given geographical location must be visible to other collections of edge nodes. 
For that, handshaking protocols and message passing are employed in discovery mechanisms~\cite{hong2019resource}.

Discovering resources and services in a distributed computing environment is a well-explored research topic, and many techniques have been employed to map tasks onto the most appropriate resources to improve performance~\cite{amento2016focusstack, liu2016paradrop, varghese2017edge}. 
FocusStack~\cite{amento2016focusstack} discovers edge nodes in a single-tenant environment whereas ParaDrop~\cite{liu2016paradrop} and Edge-as-a-Service (EaaS)~\cite{varghese2017edge} operate in multi-tenant edge environments.
However, there are additional challenges that need to be addressed to enable discovery when multiple collections of edge nodes are federated.

The FocusStack framework was proposed for the discovery of edge resources.
This framework addresses the geographical location for deploying workloads in a single-tenant environment~\cite{amento2016focusstack}. 
The authors employ it for video sharing among mobile devices.
FocusStack comprises two sub-systems, the Geocast primitive and the OpenStack.
The Geocast (Georouter) provides Location-based Situational Awareness (LSA) of edge devices to the second component, while the OpenStack extension allows the deployment, execution, and management of containers on small edge computing devices with limited networking capabilities. 
The FocusStack framework does not provide an orchestrator for edge devices.
However, this is important when the edge network has heterogeneous resources and  provides neither privacy nor security, yet there are important aspects when working with machine learning models that use edge device information.

The ParaDrop framework~\cite{liu2016paradrop} works with multi-tenant application orchestration across multiple edge nodes. 
This framework can be implemented in WiFi access points or other wireless gateways, thus allowing greater computational capacity to homes and enterprises.
"This framework is divided into three components: a flexible hosting substrate in the WiFi APs (Access Points) which supports multi-tenancy; a cloud-based back-end where computations are orchestrated across ParaDrop APs; and an Application Programming Interface (API) which can be used for the deployment of third-party developers' computing functions across different ParaDrop APs"~\cite{liu2016paradrop}.
However, this framework does not provide mechanisms for targeting federated learning at the edge or any collaborative edge computation.

The EaaS platform~\cite{varghese2017edge} is a framework that includes a lightweight discovery protocol for a collection of homogeneous edge resources. 
The EaaS platform, like Paradrop, comprises a three-tier environment:~"the top tier comprises the cloud, the bottom tier comprises user devices, and the middle tier contains edge nodes"~\cite{varghese2017edge}.
This platform requires a single master node, which can be either an available edge device or a dedicated node (edge/cloud server), to perform a management process that communicates with edge nodes.
A manager is installed on the edge nodes to execute specific commands.
These nodes are selected based on  communication with the master node.
Available administrative control panels exist on the master node to monitor individual edge nodes on the network~\cite{varghese2017edge}.
Either Docker or LXD containers are deployed when EaaS discovers an edge node.

In~\cite{DBLP:journals/jsac/ChantreF18}, the authors explore the problem of locating edge devices in NFV-based small cell networks.
Their solution employs a multi-objective optimization technique, which concentrates on the location of edge nodes from which the virtualized broadcasting service would be turned on. 

\begin{table}[!ht]
    \caption{\uppercase{Discovery Solutions in the Literature}}
    \centering
    \fontsize{8}{10}\selectfont
    \renewcommand{\arraystretch}{1.3}
    \begin{tabular}{|p{1.2cm} | p{1cm} | p{6cm}| p{6cm} |}
 \hline 
\textbf{Ref.} & \textbf{Year} & \textbf{Proposal} & \textbf{Limitation} \\
 \hline
\cite{amento2016focusstack} & 2016 & This framework addresses the geographical location for deploying workloads from the cloud in a single-tenant environment. & This proposal does not support multitenant and collaborative computation.\\\hline

\cite{liu2016paradrop} & 2016 & This proposal supports multitenant applications at the edge. Works for homogeneous resources. & This proposal does not support heterogeneous edge environments.\\\hline

\cite{khalili2016inter, wang2016evaluation} & 2016 & Proposal of message-passing which assumes that a user device can communicate with any node in a network and submit queries, and relies on simulation-based validation. & This proposal does not support heterogeneous edge environments nor does it protect the transmitted data from cyber-attacks.\\\hline

\cite{varghese2017edge} & 2017 & Proposal of a lightweight discovery protocol for a collection of homogeneous edge resources which requires a master node. & This proposal does not support heterogeneous edge environments and has security issues.\\\hline

\cite{wang2017enorm} & 2017 & Proposal of a framework uses a central edge node to communicate with the cloud and other nodes creating a graph in the format of a tree. & The handshaking in this proposal cannot be suitable for systems with intensity communication and thousand's of nodes.\\\hline

\cite{zavodovski2018edisco} & 2018 & Proposal of a discovery mechanism based on a central server for discovery of edge devices. & Need of a central node for the communication between cloud and edge servers and does not provide support for collaborative computation.\\
\hline
\end{tabular}
\label{tab:discovery}
\end{table}
This lightweight discovery protocol implies low overhead with only a few seconds for launching, starting, stopping, or terminating containers.
This low overhead confers on advantageous to EaaS~\cite{hong2019resource}.
This platform has employed the context of a single collection of edge nodes. 
The use of machine learning models at the edge for FL, however, is yet to be introduced. 
Moreover, the employment of a central server makes it difficult to detect failures and overload conditions at the edge nodes.
The authors assume that the~"edge nodes can be queried and can, via owners, be made available in a common marketplace"~\cite{varghese2017edge}.
Moreover, a central master node increases the vulnerability to security threats.

The final discovery framework to discuss here is eDisco framework, proposed in~\cite{zavodovski2018edisco}.
It allows service providers and edge devices to discover edge servers and determine an optimal algorithm for the deployment of edge resources.
This framework is divided into three steps: determination of paths to clients, determination of on-path Domain Name System (DNS) zones, and location of the edge servers~\cite{zavodovski2018edisco}.
The framework uses a central edge node  to communicate with the cloud server and other edge nodes along with a tree connection.

In~\cite{DBLP:journals/ppna/GuevaraF21}, the authors proposed an algorithm to decide where applications should be executed on fog or cloud nodes.
Their solution comprises two schedulers based on integer linear programming considering resources such as available processing capacity, memory capacity, and available storage capacity.
However, their proposal has limits to the execution of the FL since information of resources may change.

Exploiting the edge of the network requires discovery mechanisms to find suitable edge nodes to improve decentralized edge configuration, but these mechanisms need automation due to the volume of different edge devices available at the edge. 
Future research should consider the heterogeneity of edge devices from multiple generations as well as current workloads such as large-scale machine learning tasks that have not been considered in most existing work.
Benchmarking methods will need to collect information about the availability and capability of resources.
Solutions must allow for seamless integration (and removal) of edge nodes in the computational workflow at different hierarchical levels without increasing latency or compromising the user experience. 
Moreover, automatic recovery and fault handling mechanisms at the edge nodes should be adopted. 
Existing methods use cloud servers, but this is not practical in the context of artificial intelligence executing at the edge since the discovery of edge nodes not only demands large volumes of data but also data privacy.

\subsection{Deployment of Service and Application}\label{deployment}

The problem of the configuration of services and applications at the edge is generally known as the deployment problem.
In edge computing, when a collection of end-user devices require service at the edge, first, the potential edge server must be determined via discovery.
The most suitable edge node will be chosen as the edge server for execution. 
Two different scenarios must be considered for this operation: one in which the server is near to the end-users; and one in which the potential server resides in another geographic region.

In~\cite{jiang2020edge}, the authors define the deployment problem as a bi-criteria optimization problem with objectives of minimization of edge network delay and the cost of deploying computing resources. 

Generally, a service that fulfills requests from user devices must be offloaded onto one or to a set of edge nodes.
However, this is only possible if knowing the capabilities of the target edge nodes are known so that a  match of the node capabilities with the requirements of services or applications can be evaluated.

Three virtualization technologies are often used in edge deployment: Virtual Machines (VM), containers, and unikernels.
A VM provides an abstract machine that uses device drivers targeting the abstract machine, while a container provides an abstraction of operating system.
A hypervisor runs VMs that have their the operating system using hardware VM support.
Containers on a host (edge server) share the same kernel; thus, they require an underlying operating system that provides the basic services for all the containerized applications using virtual-memory support to isolation. 
The use of containers involves a lower overhead when compared with VMs, especially in environments with thousands of deployed containers. 
However, the service isolation in container systems can limit access to resources.
Unikernels are similar to containers but with their own kernel.

The Edge Node Resource Management (ENORM) framework~\cite{wang2017enorm} addresses the deployment on individual edge nodes. 
Similar to the EaaS platform, ENORM operates in a three-tier environment, but there is no single master controller for the edge nodes.
Instead, they are assumed to be visible to cloud servers and available for use by these services.
This framework makes it possible to partition a cloud server and offload it to edge nodes  to improve the overall Quality of Service (QoS) of an application.
The framework is underpinned by a provisioning mechanism for deploying workloads from a cloud server onto an edge server. 
Using handshaking protocols, cloud servers and edge servers establish connections that  can guarantee resource availability to a request executed at the edge.
The provisioning mechanism underlying this framework considers the entire lifecycle of an application server from offloading it on  edge via a container until its termination and notification of the cloud server.

A classification of services based on their Quality-of-Service (QoS) was introduced in~\cite{DBLP:conf/latincom/GuevaraBF17}.
The authors proposed a classification to help decision-making mechanisms, such as the deployment of services.
As an extension, in~\cite{DBLP:journals/jnca/GuevaraTF20}, the authors explored the classification of services, including machine learning, but their work did not explore distributed machine learning algorithms (e.g., federated learning).
A solution for location decisions for deployment in fog computing networks was explored in~\cite{DBLP:journals/sensors/SilvaF19}. 
The authors employed a multi-criteria mixed-linear programming model to handle this problem in order to improve deployment and users experience.

A special deployment framework for networks, Unmanned Aerial Vehicles
(UAVs), was proposed in~\cite{chen2017caching}.
This framework uses a machine learning technique to predict user behavior, but the authors did not explore federated learning.

In~\cite{6600983}, the authors explored the problem of content deployment in small base station networks.
Their solution to alleviate the bottleneck was to employ  multiple caches to reduce the  downloading delay for mobile users.

Huawei developed an open-source edge computing framework based on Kubernetes for networking called KubeEdge \cite{kubeedge}.
Kubernetes is an orchestrator of containers that allows automating computer application deployment, scaling, and management.
However, Kubernetes would impose limitations for FL since it does not provide continuous deployment of the containers, load balancing for multiple services, scaling, and the necessary security.
Since FL continuously aggregates the model results of edge devices and updates the models, this would be a problem for Kubernets as well as managing the load balance across the edge.
Moreover, the solution in \cite{kubeedge} neither supports nor guarantees the isolation of multiple services and applications executed at the edge.

Two frameworks for deployment in the Internet of Things (IoT) have been proposed, Azure IoT Edge~\cite{azure} and EdgeX~\cite{edgex}, which were devised for delivering cloud intelligence to the edge by deploying and running artificial intelligence on cross-platform IoT devices.
Azure IoT Edge~\cite{azure} allows machine learning algorithms to be executed at the edge and uses containers for deployment.
This module creates an IoT hub, registers an IoT Edge device on that IoT hub, and installs and starts the IoT Edge runtime on a virtual device.
Remotely, the Azure IoT Edge deploys a module to an IoT Edge device and sends telemetry to IoT Hub.
The EdgeX~\cite{edgex} is a vendor-neutral open source project focusing on Industrial IoT Edge.
This framework also allows machine learning algorithms to be executed.
For deployment, the EdgeX uses microservices with plug-and-play components to unify the marketplace and accelerate the deployment of IoT solutions.

\begin{table}[!ht]
    \caption{\uppercase{Deployment Solutions in the Literature}}
    \centering
    \fontsize{8}{10}\selectfont
    \renewcommand{\arraystretch}{1.3}
    \begin{tabular}{|p{1cm} | p{1.3cm} | p{6cm}| p{5.4cm} |}
    \hline
 \textbf{Ref.} & \textbf{Year} & \textbf{Proposal} & \textbf{Limitation} \\
 \hline

\cite{talagala2018eco} & 2018 & Orchestrator to manage machine learning deployments including distributed learning such as FL. & Flexibility and scalability issues as well as privacy concerns.\\\hline

\cite{7356560} & 2015 &  DDF programming model in the context of fog computing; the model based on the MQTT protocol supports the deployment of flow on multiple nodes, and assumes the heterogeneity of devices. & No support of FL nor anomaly detection during deployment. In order to support FL it also needs to support heterogeneity of data and communication resources.\\\hline

\cite{fate2019, ludwig2020ibm} & 2019-2020  & Platforms developed specifically for the deployment of FL algorithms. & Limited number of FL techniques. Limited number of machine learning algorithms and they do not provide resource management.\\\hline

\cite{varghese2017edge} & 2017 & Operation in a tree-tier environment. Deploying applications via Docker or LXD containers. & Does not support collaborative edge. The major drawback is the centralized master node.\\\hline

\cite{wang2017enorm} & 2017 & Assumption that edge nodes are visible to cloud servers that may want to make use of the edge. The framework allowing for partitioning a cloud server and offloading tasks to edge nodes for improving the overall QoS of the application. Use of containers for deployment. & Does not support collaborative edge computation. This makes FL impossible since edge devices must collaboratively execute model training be coordinated by edge/cloud server.\\\hline

\cite{hosseinalipour2020multi} & 2020 & Horizontal federated aggregation based on a distributed average consensus formation scheme. Device energy should safe and the also reduction of the number of parameters transferred over the network compared to conventional FL. & This research only explores the aggregation, no consideration of energy consumption, time, and learning. Moreover, these solution focus only on aggregation and not on the other steps that include edge devices selection, resource allocation, necessary for FL at the edge.\\\hline

\cite{ye2020federated} & 2020 & Federated aggregation using $FedAvg$ which is a synchronous technique that updates the central model in each round, server must be updated before edge devices. & Synchronous techniques not suitable for most of real-world scenarios with heterogeneity of resources at the edge (e.g. CPU, memory, battery) and resource communication (e.g. WiFi, 4G, 5G).\\
\hline
\end{tabular}
\label{tab:deployment}
\end{table}
In~\cite{fate2019}, the authors propose an open-source framework called FATE.
FATE is an industrial-level FL framework, which aims at providing FL services between different organizations. 
It supports training various machine learning models under both horizontal and vertical federated settings, and it integrates secure multi-party computation and homomorphic encryption to provide privacy guarantees. 
The FATE-Flow platform helps users define the pipeline of their FL process, including data pre-processing, federated training, federated evaluation, model management, and model publishing. 
FATE-Serving provides  inference services to the users.
This component supports loading FL models and conducting online inference.
FATE has a visualization tool that provides visual tracking of the job execution and model performance, and it uses KubeFATE to help deploy FATE on clusters through the use of Docker or Kubernetes.
It provides customized deployment and cluster management services. 
In the FATE framework, practitioners must modify the source code in order to create a new federated algorithm, although this procedure is not easy for non-expert users~\cite{li2019survey}.

The IBM federated learning framework was introduced in~\cite{ludwig2020ibm}.
This framework is based on two components, parties (edge devices) and an aggregator.
The IBM framework focuses on~"secure deployment, failure tolerance, and fast model specification"~\cite{ludwig2020ibm}.
One advantage of this framework is the allowing connection types, including the Flask web framework, gRPC, and WebSockets.

A deployment strategy specific for machine learning was introduced in~\cite{talagala2018eco}, where an orchestrator manages the deployment of learning models.
This orchestrator also supports federated learning algorithms provided by the Tensorflow framework, but there are still limits in terms of scalability and flexibility.
Another framework is the NVIDIA Clara~\cite{clara2019}, a Software as a Service (SaaS) for healthcare and hospitals.
This framework can work with FL, especially in distributed smart embedded applications.
For deployment, Clara includes the necessary building blocks and reference applications that enable developers to build clinical workflows. 
However, the framework limits access and makes necessary the use of specific technologies and hardware provided by NVIDIA. 
Clara also employs containers and Kubernetes for container orchestration. 

The deployment decision on where training, inference, and aggregation will be executed must be carefully made.
Most existing research considers that federated aggregation will be executed in a central server (cloud)~\cite{nishio2019client, fate2019, li2019fair, wang2020attention, lin2021deploying}, but some edge servers/devices are unsuitable for the execution of  such task because of the resource limitation (e.g., communication technology of edge devices).
To handle this problem, in~\cite{hosseinalipour2020multi}, the authors proposed a horizontal federated aggregation based on a distributed average consensus formation scheme at the edge. 
The authors showed that even with limited Device-to-Device (D2D) communications enabled, the learning accuracy achieved was comparable to that of the centralized gradient descent algorithm.
The results demonstrated that device energy could be saved and the number of parameters transferred over the network when compared to traditional federated learning.

\begin{table}[!ht]
    \caption{\uppercase{Load Balancing Solutions in the Literature}}
    \centering
    \fontsize{8}{10}\selectfont
    \renewcommand{\arraystretch}{1.3}
    \begin{tabular}{|p{1cm} | p{1cm} | p{6cm}| p{6cm} |}
    \hline
     \textbf{Ref.} & \textbf{Year} & \textbf{Proposal} & \textbf{Limitation} \\\hline
     \cite{mtibaa2013towards} & 2013 & Focuses on the balancing of power consumption, where edge devices decides whether to offload tasks considering the power of edge batteries. & Not suitable for use in scenarios with FL at the edge.\\\hline
     \cite{hung2018videoedge} & 2018 & Specific for video processing at the edge. & Does not support a variety of data as well as FL at the edge.\\\hline
     \cite{dai2018joint} & 2018 & Optimization of the offloading for VEC. & No support for dynamic scenarios.\\\hline
     \cite{cho2020client} & 2020 & Client selection algorithm for federated learning. & Focuses only on the performance of training and only employs federated averaging for aggregation. \\\hline
     \cite{duan2019astraea} & 2019 & Creation of balance at the edge employing a machine learning solution.& No solution for imbalance of workload.\\
     \hline
    \end{tabular}
\label{tab:load_balancing}
\end{table}
The authors in~\cite{ye2020federated} explored federated aggregation in vehicular edge computing using $FedAvg$ in a central aggregator.
Their solution explored problems related to the diversity of image quality and computation capability for vehicular clients.
To select nodes for aggregation, they employed Deep Neural Networks (DNN) at the edge, creating contracts between a central node and vehicular clients when they are suitable for aggregation.
The selection of a local DNN model with suitable image quality and computation capability was formulated as a two-dimensional image-computation-reward contract-theoretic problem.
Their approach does not explore the heterogeneity of the network at different levels (hardware, software, and data), even though this impacts  decisions about the FL cycle of training.
Moreover, their proposal did not explore the whole process of deployment of FL models at the edge.

\subsection{Resource Allocation}\label{resource-allocation}

Resource allocation refers to the process of distributing tasks over a set of resources so that the overall processing will be more efficient, i.e., the distribution of tasks tries to optimize a given objective function. This is challenging mainly when edge servers and edge devices are distributed geographically and have different hardware and software configurations.
Each edge server can serve a varying set of edge devices, which helps to reduce the complexity of monitoring and resource management.
Tasks can be performed on one or more edge servers and executed in parallel on the cloud.
The execution of tasks can be split among the edge and cloud servers to overcome edge resource limitations, but this makes the management of resources more complex as there are three entities with different characteristics that need to be managed (cloud servers, edge servers, and edge devices). 
Deciding where tasks will be executed  to optimize an objective function makes  resource management challenging, since mobility, computation, and communication must be considered.

One common objective function in resource allocation is load balancing, where tasks are distributed evenly among servers to avoid overwhelming their computing capacity and, thus, balance the load. This objective function inherently avoids increased response times for applications due to resource overloading. Load balancing can be accomplished by implementing an auto-scaling algorithm, which dynamically adds or removes resources to cope with fluctuations in the workload at the edge.
An edge node can be a traffic routing node, such as a router or mobile base station, so an offloaded service should not compromise the QoS of the essential service executed on that node. 
Edge servers are monitored in terms of  network and system performance, and an estimation is made about whether the QoS requirements can be met. 
If an edge server cannot provide the QoS requirements for a task, then the resources allocated for that application are scaled.
Most load balancing algorithms in the literature do not provide solutions to balance the load resulting from FL processing.

The authors in~\cite{mtibaa2013towards} have proposed a load balancing algorithm for mobile edge computing, which focuses on balancing the energy consumption among the edge devices, with offloaded tasks considering the state of the batteries.
However, this algorithm does not support FL, and balance might jeopardize the collaborative interaction of the FL model.

Monitoring edge resources is a crucial requirement for the effectiveness of load balancing as well as the achievement of auto-scaling methods.
Existing monitoring systems for distributed systems do not scale and yet consume  considerable  resources; consequently, they are not suitable for large-scale resource-constrained edge deployments. 
Current mechanisms for auto-scaling resources are limited to single-edge nodes and use lightweight monitoring~\cite{wang2017enorm}. 

\begin{figure}[!ht]
    \centering
    \includegraphics[scale=0.4]{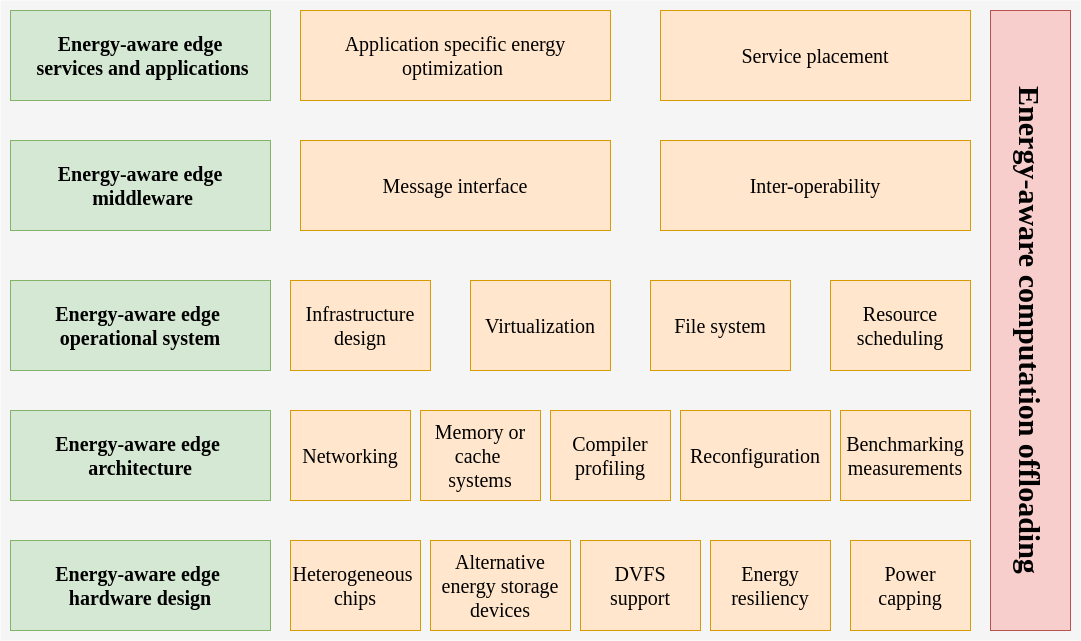}
    \caption{Complete view of energy awareness in edge computing~\cite{yang2019energy}.}
    \label{fig:energy-aware}
\end{figure}
 
In~\cite{hung2018videoedge}, the authors propose an end-edge-cloud hierarchical architecture, called VideoEdge, to help achieve load balancing in analytical tasks while maintaining high accuracy.
The VideoEdge is specific for video processing, executing tasks at the edge, which would generally demand high bandwidth to send to the cloud.
In their proposed network architecture, cameras are connected to clusters (sets of edge servers). 
Their architecture identifies the most promising \textit{configurations} and filters out those with  low accuracy and great  resource demands.
However, the work does not explore FL for training models and cannot work with different data formats, which is a necessary feature as multiple edge devices in the real world can exist in the same local area.
One example would be an industry that has cameras, vehicles, robots, embedded sensors, and machines, generating different formats with/without mobility, resulting in highly complex heterogeneous scenarios for balancing the workload on edge nodes.

The authors in~\cite{dai2018joint} propose a load balancing solution for Vehicular Edge Computing (VEC), which optimizes the computation offloading strategy and resource allocation while considering the mobility of vehicles as part of the load balancing problem.
The authors define a utility function for balancing the load among VEC servers and maximizing this function. 
However, their solution cannot be applied in dynamic scenarios with cars arriving and departing on the one-way unidirectional road.
Since federated aggregation model training may run on edge servers/devices, the mobility of vehicles can force the migration of tasks, which can increase the difficulty of creating a balanced load at the edge.
Moreover, the solution proposed does not work with artificial intelligence algorithms, which increases the complexity of balancing load since these algorithms work only on data sets.

The size of datasets on the edge devices can vary, which can introduce biases in training results and reduce the performance of the FL model executed by the FL.
This has motivated the authors to introduce an algorithm in~\cite{duan2019astraea} that tries to promote balance using machine learning techniques, but their solution focuses only on the data imbalance. It does not provide a solution for the workload imbalance resulting from the parallel execution of tasks by edge devices/servers.
A client selection algorithm for training models focused on load balancing is introduced in~\cite{cho2020client}, based on the assumption that biased client selection affects model convergence.

\subsection{Energy Consumption}\label{energy}

The energy consumed by machine learning algorithms in data centers has been gaining attention 
lately~\cite{dayarathna2015data} since it can  considerably impacts  Power Usage Effectiveness (PUE).
However, edge devices and servers do not usually have the special air refrigeration that data centers do, and this lack of support can impact  the execution of FL at the edge.
So, the quality of services using network edge is still an open problem~\cite{8057196} since we need to provide a balance between resource usage and learning results as well as to deliver a service requirement in a short span of time.
For example, autonomous navigation systems use a lot of data and demand great energy consumption; moreover, existing processors consume a large amount of energy, which limits the execution of Deep Learning services at the edge.

Equation~\ref{eq:energy-eq} gives the energy consumed by a device service using cloud computing~\cite{zhang2017energy}.
\begin{table}[!ht]
    \caption{\uppercase{Proposals for Energy Efficiency in the Literature}}
    \centering
    \fontsize{8}{10}\selectfont
    \renewcommand{\arraystretch}{1.3}
    \begin{tabular}{|p{1.4cm} | p{1.3cm} | p{5.6cm}| p{5.6cm} |}
    \hline
\textbf{ Ref.} & \textbf{Year} & \textbf{Proposal} & \textbf{Limitation} \\\hline
 \cite{zhang2016energy, zhang2017energy} & 2016-2017 & Mechanisms to minimize energy by the use of offloading. & No exploration of energy consumption in collaborative scenarios with the communication between edge devices.\\\hline
 \cite{wang2019adaptive} & 2019 & Proposal to adapt both bandwidth and computation resources based in order to reduce energy consumption. & Not suitable for heterogeneous environments.\\\hline
 \cite{yang2020energy, zaw2021energy} & 2020-2021 & Proposal focusing on energy consumption to facilitate FL. & Focused only on energy consumption of edge devices.\\\hline
 \cite{li2020talk} & 2020 & Proposal for a compression mechanism to balance the energy consumption of edge devices for wireless networks. & Considers only edge devices as the energy consumer, but federated learning may involve cloud and edge servers.\\\hline
 \cite{zaw2021decentralized} & 2021 & Proposal for a mechanism for energy-aware resource management for FL using a decentralized approach and game theory. & Increased complexity as the number of edge devices increases.\\\hline
\end{tabular}
\label{tab:energy}
\end{table}

It takes into consideration the energy consumed by device gateways when receiving data from devices and sensors ($E_{GW-r}$), the energy consumed by device gateways to transmit data to the cloud data center ($E_{GW-t}$), the energy consumed by the transport network between device gateways and cloud ($E_{net}$), and the energy consumed by components of the data center ($E_{DC}$). 

\begin{equation}\label{eq:energy-eq}
    E_{dev-cloud} = E_{GW-r} + E_{GW-t} + E_{net} + E_{DC}
\end{equation}

\begin{equation}\label{eq:energy-dev}
    E_{dev-edge} = E_{GW-r} + E_{GW-c} + \beta(E_{GW-t}+E_{net}+E_{DC}).
\end{equation}

Equation~\ref{eq:energy-dev} can be used to calculate the energy consumption of communication between the device and the edge. 
This equation also takes into account $E_{GW-c}$, the energy consumed by device gateways for local computation and processing, and the ratio of the number of updates from the edge to the cloud for synchronization. 
Mechanisms at the edge should consider all these variables and control the number of updates from the edge to the cloud to reduce energy consumption while keeping learning accuracy at acceptable levels. 
Such trade-offs remain an open problem yet to be addressed.

Several tasks conducted in FL, such as data processing, resource management, transmission optimization, and routing, need to consider the energy consumption. 
However, limited resources in the edge devices, such as limited storage and  computing power, prevent the adoption of more elaborate management approaches on mobile devices and edge nodes~\cite{wang2020convergence}.

Most existing work aimed at minimizing energy consumption focus on edge offloading~\cite{zhang2016energy, zhang2017energy}.
In~\cite{zhang2016energy}, the authors study the energy consumption of Mobile Edge Computing (MEC) in 5G heterogeneous networks.
Their strategy decreases the energy consumption using an energy-efficient computation offloading mechanism under latency constraints of the computation tasks.
The sharing of edge offloading can transfer computation from a resource-limited mobile device to a resource-rich MEC server to improve the execution of mobile applications.
Each mobile device can decide whether to offload its task for remote computing or to execute a task locally on the device, using classification and priority assignment.
However, this strategy does not explore energy consumption in FL, in which the communication between edge devices and MEC servers occurs frequently. 
Training models running on edge devices decrease  communications but still consume a considerable amount of energy.

The authors in~\cite{zhang2017energy} propose a solution to balance the energy consumption and latency in mobile edge computing.
Their solution optimizes communication and computational resource allocation under limited energy and sensitive latency. 
They use an iterative search algorithm that simplifies the problem and reduces the computational complexity.
However, their simulations do not consider the execution of tasks that require neither learning algorithms nor FL to manage the energy 
consumption across the edge.
The studies in~\cite{zhang2016energy} and \cite{zhang2017energy} focus on offloading, but they do not try to minimize the energy consumed in the computation of tasks at the edge.

In~\cite{yang2019energy}, the authors surveyed multiple proposals for reducing the energy consumption in edge computing (Fig~\ref{fig:energy-aware}).
Their survey explores the hardware layer to the application layer, including heterogeneous architectures, energy storage devices, energy resiliency, power capping, and Dynamic Voltage and Frequency Scaling (DVFS) support.

The authors in~\cite{wang2019adaptive} focus on achieving optimal training of learning models; they propose a solution that adapts both bandwidth and computation resources based on the number of local iterations at each device as well as the number of global iterations.
However, no efficient solutions for heterogeneous environments are provided.

In~\cite{kobayashi2019radio}, the authors introduce a proposal to reduce energy consumption in multi-access edge computing.
Their proposal optimizes the allocation of radio and computational resources to minimize the total processing time and reduce energy consumption, but the federated learning paradigm was not considered.

A solution for energy-efficient radio resource management is proposed by exploring bandwidth allocation and scheduling~\cite{zeng2020energy}.
The solution tries to optimize resource utilization while providing learning guarantees, such as the maximization of the number of collected updates and maintenance of the level of local accuracy.

In~\cite{zaw2021energy}, the authors propose an energy-aware solution for resource management in federated edge computing. 
This work considers multi-access edge computing as the edge environment, and the proposal tries to improve training model performance, reduce the total time required to perform model training, and energy consumption of mobile devices by offloading part of the local dataset to the edge server to decrease the energy consumption of edge devices.

Most of the existing proposals do not explore federated edge computing, where tasks are executed collaboratively, so the balance between splitting a task across an edge and energy consumption remains as a challenge.

\subsection{Migration}\label{migration}

In edge computing, migration involves transferring a task from one edge node (server or device) to another one that has the capability to execute it.
Migration is crucial for task execution under certain circumstances such as hardware failure, network overload, user mobility, or power consumption~\cite{7424534}.

When employing federated learning, the generated data will not be sent through the network to other edge nodes, since each node is responsible for carrying out model training with its  data, so the migration should not happen during model training, but it can happen for other processes, such as aggregation on edge servers.
Since migration can still happen, secure migration mechanisms need to be developed to guarantee  security during the migration of services across edge servers~\cite{yu2020deep}.

Mechanisms for migration are often necessary to guarantee task continuity.
As users move farther away from the edge server executing a task, it must then be delegated to another server, however, without interrupting the task.
According to~\cite{wang2018survey}, the migration of tasks must deal with decisions about whether the ongoing task should be migrated out of the current edge server; and the selection of another edge node to execute such that task and how migration should be carried out considering overhead and QoS requirements.
Moreover, the migration across edge nodes requires information about network connectivity, disk  and RAM state, among others.

\begin{figure}[!ht]
    \centering
    \includegraphics[scale=0.4]{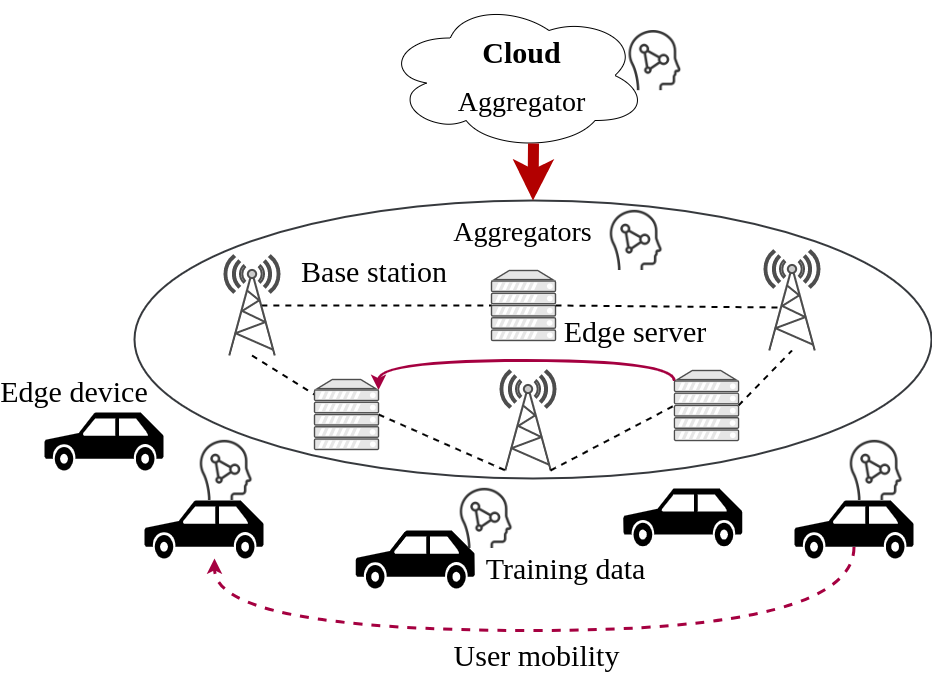}
    \caption{A case of migration in mobile edge computing. The solid red line exemplifies a single path of one transfer between source and destination edge server~\cite{wang2018survey}.}
    \label{fig:migration}
\end{figure}

When migrating a virtual server the traffic volume cannot be neglected because of the large size of stored information. 
The cost of migrating a virtual server depends on the size of the server and the bandwidth available on the migration path. 
According to~\cite{chen2019dynamic},~"For example, for an emotion detection service, the migration cost depends on the emotion recognition models.
In video-streaming services, the migration cost depends on the data size of the decoded video".
In~\cite{chen2019dynamic}, the authors model the problem of migration for edge computing and conclude that a higher service resolution results in greater migration costs.

Migration of virtual machines (VMs) across clusters is possible, but in practice, it results in  significant time overhead. 
Moreover, live migration between geographically distributed cloud data centers is very challenging and time-consuming since tasks must move across servers without disconnecting user devices with the applications.
Similar strategies have been adopted in edge resources for live migration of VMs~\cite{callegati2013live, darsena2013live}.

Even though it is possible, the migration can take a long time, making it challenging to use existing strategies for real-time, interactive applications.
Additionally, VMs cannot be standardized for hosting tasks at the edge~\cite{shi2016edge, shekhar2017dynamic}. 
Containers can be a lightweight alternative technology.
Strategies involving migration need to incorporate container technologies.

In~\cite{gomes2017edge}, the authors introduce a content-relocation algorithm for migrating cached content in edge devices.
Their solution employs a prediction mechanism for monitoring the mobility of edge devices to perform the migration. 
However, the paper does not consider FL, where edge devices can train models and send data to an edge/cloud server, making it more complex due to the limits of data sharing, resources, and location.
Since edge servers can execute federated aggregation, migrating this task can be complex due to the location as well as the resulting overhead.

Figure~\ref{fig:migration} shows an example of migration in mobile edge computing.
In this example, an edge server contains one or more physical machines hosting several virtual machines and is able to cover the nearby mobile users~\cite{wang2018survey}. 
These edge servers can be interconnected with each other using different kinds of network connection technologies. 
In~\cite{wang2018survey}, the migration is considered a stateful migration of applications, where the acceptance of tasks by mobile edge devices are continuous, as well as the resource reservation of internal state data for edge devices.
In their solution, the task resumes where it stopped before migration~\cite{wang2018survey}. 
As a mobile edge device moves from one area to another, there are two options, either the task will continue running on the current edge server and exchange data with a mobile edge device through the core network or other edge servers, or it will migrate to another edge server that covers the new area~\cite{wang2018survey}.
In either case, the cost will be that of the data transmission for the former case, and that of the migration for the entire task for the latter.

In~\cite{saurez2016incremental}, the authors introduced the Foglets platform, which uses mobile agents to allow migration.
The runtime state of a task is captured at a high level by the application itself and then migrated to a new edge node.
The solution proposed captures the execution state with only a coarse granularity, as it does not allow the destination node to restore the state of a thread at the exact instant of checkpointing.
The authors in~\cite{bellavista2017migration} present a platform capable of proactively migrating the entire runtime state of a task by actually migrating VMs. 
More specifically, their proposal extends the Openstack++11 platform to enable mobility support.
In~\cite{bao2017follow}, the authors considered situations in which a single edge node is given a specific access point.
Moreover, parameters such as the state of hardware resources are not included in their migration decision-making.

\begin{table}[!ht]
    \caption{\uppercase{Migration Solutions in the Literature}}
    \centering
    \fontsize{8}{10}\selectfont
    \renewcommand{\arraystretch}{1.3}
    \begin{tabular}{|p{1cm} | p{1cm} | p{5.8cm}| p{6cm} |}
    \hline
 \textbf{Ref.} & \textbf{Year} & \textbf{Solution} & \textbf{Limitation} \\
 \hline
 \cite{callegati2013live, darsena2013live} & 2013 & This solution explores live migration of virtual machines. & This solution is not suitable for federated learning. \\\hline
 \cite{bao2017follow} & 2017 & This solution is suitable for scenarios with only a single edge node connected to an access point. & Consideration of limited number of parameters to perform migration.\\\hline
 \cite{wang2018survey} & 2018 & A survey of migration exploring mobile edge network.& Does not explore migration of services that use FL. \\\hline
 \cite{bellavista2017migration} & 2017 & Proactive migration employing VM for the migration procedure. & Does not provide a solution to migrate tasks such as federated aggregation.\\\hline
 \cite{saurez2016incremental} & 2016 & This solution uses the mobile agents to allow migration. & No provisioning for the destination node to restore the state of a thread at the exact instant of check pointing.\\\hline
\cite{chen2019dynamic} & 2019 & A cognitive architecture for migration for IoT networks, this proposal tries to automate the migration by using learning algorithms. & It works only for IoT applications. It has not been test for FL applications.\\ 
\hline
\end{tabular}
\label{tab:migration}
\end{table}

In~\cite{chen2019dynamic}, the authors introduce a cognitive edge computing architecture for dynamic migration.
Their proposal focuses on IoT applications that execute artificial intelligence algorithms using data provided by video streaming.
In their proposal, $Q$-learning is used to make decisions based on information and migration strategy.
However, the proposal neither supports service migration in federated edge computing nor does it consider tasks in the event of failures or load balancing.

\section{Challenges and Future Directions}\label{challenges}

In this section, we highlight some of the challenges facing edge computing and edge intelligence, federated learning,  resource management at the edge, and future directions.

\subsection{Challenges in Edge Computing and Edge Intelligence}

In~\cite{wang2020convergence}, the authors describe issues to be faced when extending deep learning from the cloud to the edge of the network.
Several constraints of networking, communication, computing power, energy consumption, and devising the development of edge computing architecture need to be considered to achieve the best performance of training models and inference. 
As the computing power of the edge increases, edge intelligence will become common, and the intelligent edge will play an important supporting role in improving its performance and efficiency.
The authors also discuss caching policies, improvement of edge architecture, deployment, and how to modify training to enhance the performance of learning algorithms, especially Deep Learning (DL).

Issues in resource management that require solutions include balancing the workload on edge and server devices, balancing interruption requests between storage servers, migration of services and applications across the edge, and distribution and allocation of tasks on the edge. 
Moreover, another challenge will be to ensure task completion, given delay constraints.
Optimization of the use of resources on the edge should be able to execute various deep learning services, as well as guarantee QoS requirements.

The distributed nature of edge computing and the heterogeneity of edge nodes permeate the major challenges for edge intelligence.
Different edge devices and servers are distributed across different geographical regions, and they can run different AI models and tasks.
It is thus important to design efficient service discovery protocols and to fully exploit the resource across edge nodes.
Complex edge AI models need to be partitioned into small sub-tasks, and the offloading of tasks for collaborative execution must be carried out efficiently.

Promising applications to be hosted on intelligent edge include intelligent driving, gaming, virtual reality, healthcare, and industry 4.0. 
Various machine learning applications in intelligent driving, such as collision warning, require edge computing platforms to ensure millisecond-level interaction delay. 
Moreover, edge perception is more conducive to the analyses of the immediate traffic environment surrounding the vehicle, which should enhance driving safety~\cite{wang2020convergence}.
For healthcare, assemble machine learning based on a smart healthcare system for autonomic diagnosis can execute tasks at the edge nodes.
The integration of Internet-of-Things (IoT) devices and edge computing can be challenging as it calls for data protection, constant communication, as well as fast data processing, given the resource limitation at the edge.

\subsection{Challenges in Federated learning}

Challenges in the operation of federated learning need to be incorporated into mechanisms of resource management to optimize the performance of the network, computational resources, and model accuracy.
The following challenges should be the focus of attention:
\begin{figure}[!ht]
    \centering
    \includegraphics[scale=0.18]{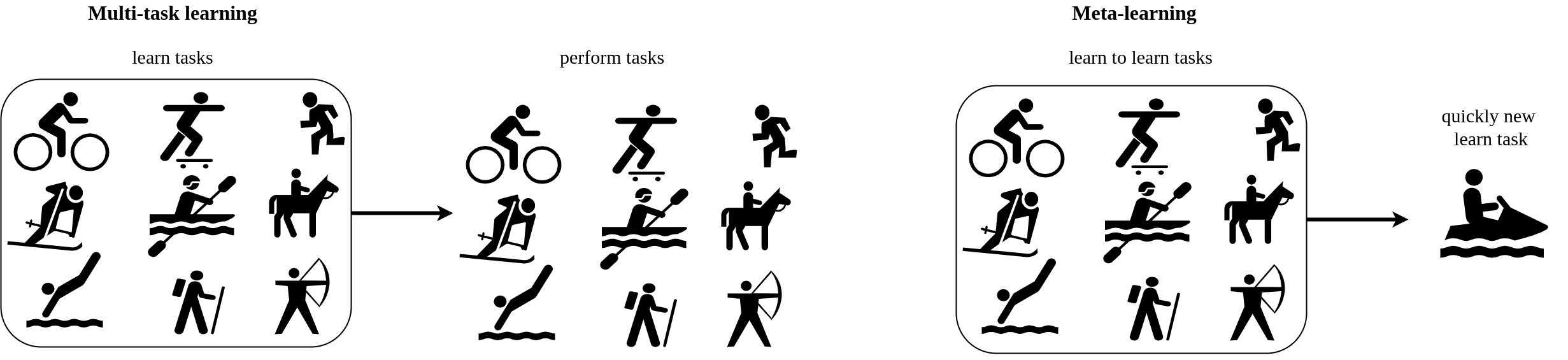}
    \caption{The difference between multi-task learning and meta-learning frameworks using Reinforcement Learning~\cite{metalearning}.}
    \label{fig:mtvsml}
\end{figure}
\begin{itemize}
    \item Heterogeneity of local datasets, which may result in bias related to population distribution. Moreover, the size of the datasets may vary significantly. For resource management solutions, heterogeneity impacts  the selection of edge devices to train the learning model because of the time to execute as well as the global result of training.
    
    \item Variations in local dataset over time due to different hardware (CPU and memory), network connectivity (e.g., 3G, 4G, 5G, or Wi-Fi), and power (battery level). This has a similar impact to the data heterogeneity as it impacts  data availability, also impacting time to convert a  model and the cost of training.
    
    \item Interoperability of node datasets is a prerequisite, potentially requiring regular curation. Management mechanisms can be developed to provide a curation.
    
    \item Federated learning makes a step toward protecting data generated on each device by sharing model updates, e.g., gradient information, rather than the raw data. However, hiding training data might allow attackers to inject backdoors into the global model. Resource management solutions need to work together with security and privacy approaches to improve the processing of management (e.g., edge node selection in each round).
    
    \item Lack of access to global training data makes identification of undesirable biases. Resource management can help in selecting devices at different places to aggregate data properly to avoid biases.
\end{itemize}

The authors in~\cite{bonawitz2019towards} discuss that aggregation methods, such as federated averaging, are unsuitable for parallel systems with thousands of devices since it can result in inefficient model convergence.
New algorithms must consider parallelism in large-scale systems.
One of the challenges in executing FL at the edge~\cite{smith2017federated} is related to statistical representativeness.
One possible solution is multitask learning and meta-learning frameworks, which allow the training of more than one model~\cite{smith2017federated, chen2018federated, zhang2017survey, fallah2020personalized, wu2020personalized}.

In multi-task learning and meta-learning, the goal is to learn a set of skills. 
Multi-task Reinforcement Learning (RL), illustrated in Fig~\ref{fig:mtvsml}, assumes that one wants to learn a fixed set of skills from minimal data, while in meta-RL, the goal is to get experience from a set of skills to learn to solve new skills quickly~\cite{metalearning}.

The incorporation of meta-learning and multi-task in FL is challenging.
One of the initial challenges is the difficulty of determining whether the FL even needs to incorporate them.
The implementation of meta-learning and multi-task learning is more time-consuming than traditional techniques, and specialists are needed to design suitable models.
Another challenge specific to multi-task learning is knowing what to share, requiring understanding the problem that must be dealt with.
Moreover, with more features (in the case of supervised models), computational processing increases, potentially compromising the execution at the edge since FL cannot share raw data.
Strategies to handle such problems should be explored in the future. 

Heterogeneity of data is another problem for edge resources and communication technology. Personalized learning models for edge devices may be a solution to provide a balance between resource usage and learning results.
In~\cite{kulkarni2020survey, sim2019investigation, tuor2020data}, the authors discuss possible techniques to personalize learning models to improve the performance of federated learning at the edge.

\subsection{Challenges in Resource Management}

\subsubsection*{Discovery}

Edge resource discovery protocols should consider the requirements of FL models for execution on the edge. 
Since FL iteratively involves a set of edge nodes, constraints due to the diversity of requirements and the mobility of edge nodes need to be considered.

The discovery of edge devices requires the development of various mechanisms, including a handshaking protocol to establish communication with user devices and cloud servers. Moreover, communication overhead should be avoided to assure the low latency that is key to guaranteeing the Quality-of-Service (QoS).

Discovering edge resources becomes especially challenging when clusters of edge nodes need to be federated from different geographic locations to create a global architecture.
Moreover, the mobility of edge nodes makes discovery problems much harder to solve.
Mobile edge nodes over a complex networking layer cannot be relied to be present at a given moment, so the consumer service requires a real-time mechanism to locate the provider service.

\subsubsection*{Deployment}

Most deployment solutions have been developed for a centralized environment, making them unsuitable for distributed environments, such as those assumed in FL. 

Federating edge resources brings about many  scalability challenges, such as when global synchronization between different administrative domains must be maintained in a federated edge. 
Moreover, individual edge deployments will most likely have different settings, such as the number of services hosted and the number of end-users in the coverage area. 
In a federated setting different service offloads from multiple domains should be possible and will require synchronization across the federated deployments.
Moreover, as multiple services and applications can be executed at the edge, guaranteeing isolation between them is a real challenge.

When working in a collaborative scenario, it is expected that failures can occur in some edge nodes or in the communication channel.
It is thus necessary to develop strategies to handle failures, although this has had little research interest.

Benchmarking multiple edge nodes (or  collections) Simultaneously will be essential to meet the service objectives. 
This is a challenging task and must be performed in real-time.
For realistic and practical approaches to edge computing deployments, it is crucial to provide flexibility to support the operations of a federated setting in a global context.

\subsubsection*{Energy Efficiency}

Energy-aware edge computing has been widely investigated in different application domains.
Most of the existing work focuses on achieving a single objective, such as low latency, or data privacy, or power saving, or energy efficiency.
Therefore, there are several opportunities to investigate multiple objectives in  optimizing  the  main metrics, such as energy efficiency and latency. 
For example, researchers have proposed novel architectures and middlewares to provide interoperability between different edge devices and resources in edge computing. 
However, energy awareness of operating systems in edge computing is still a challenge and an open  research question. 
Moreover, efficient energy management  in heterogeneous hardware for systematic energy reduction in edge computing is also an open question. 
Currently, there are no usable benchmarks to evaluate the energy efficiency of heterogeneous edge computing architectures.

For the FL paradigm and training models, studies are needed to determine how the workload at the edge can be split into small pieces to reduce energy consumption and improve load balancing.
Mechanisms that monitor the energy consumption to answer the question of where the energy is going to and to create policies for resource allocation on the edge would be a promising way to increase energy efficiency.

 \subsubsection*{Load Balancing}

A significant number of services subscriptions  at the edge require efficient management to allocate resources for individual or collective services.
This will require adequate monitoring of resources at the edge, but traditional methods cannot be used given the resource constraints of edge nodes. 
Similarly, mechanisms will need to be put in place for scaling the resources for one service (which may be heavily subscribed) while de-allocating resources from less utilized services.
Both monitoring and scaling will need to ensure integrity so that the workload is fairly balanced.

Future work includes investigating fine-grained resource allocation/deallocation for load balancing in auto-scaling schemes. 
Current auto-scaling methods add or remove  predefined discrete units of resources on the edge.
However, this type of solution can be limited in resource-constrained environments in which resources may be over-provisioned.
Alternative mechanisms which are capable of evaluating the quantities of resources to be allocated/deallocated on the basis of specific application requirements to meet Service-Level Objectives (SLOs), but without compromising the stability of the edge environment, are needed.

Balancing performance and computation in the execution of machine learning algorithms at the edge is also a challenge.
Since supervised machine learning algorithms need data for training, and  real-world models need to be retrained constantly, it is advisable to create parameters to define when to stop training and when to retrain to reduce computation.
Approaches that provide a trade-off between load balancing and the reduction of resource consumption should be studied since this can potentially allow running more applications and services using machine learning algorithms at the edge.

\subsubsection*{Migration}

Currently, edge-based deployments assume that services running on an edge node can  be
cloned or made available on an alternative edge node.
When and how to conduct such migration are of great concern in dynamic service migration mechanisms. 
Most migration mechanisms decide when to migrate by relying only on network conditions; few of them take user behavior into account. 
However, deciding when to migrate according to user behavior and mobility could tremendously improve the user experience and resource utilization.

Ensuring reliability and service mobility is also challenging. 
User devices and edge nodes may connect and disconnect from the Internet. 
This might create an unreliable environment.
A casual end-user device will expect seamless service, perhaps via a plug-and-play functionality to obtain services from the edge, but an unreliable network could cause latency. 
In a cellular system, mobile users continuously move from one cell to another while communicating, so handover (or handoff) must be performed carefully to avoid service interruption.
The challenge here is to mitigate service interruption and create a reliable environment that supports federated edge computing. 
One mechanism for  implementing  reliability is the replication of services or the facilitation of migration of services from one node to another. 
The main challenge is to keep  overhead to a minimum so that the  QoS of an application is not affected.

\subsection*{Energy consumption}

Energy consumption in task migration is not negligible and brings additional challenges.
It's necessary to develop an energy management mechanism specific to migration.
Managing energy consumption for migration becomes more challenging as the edge nodes in the real-world are heterogeneous, so the mechanism could have to be adaptable to work with edge nodes resources.
For federated edges, the migration must establish a route in the geographical area and know which nodes in the federated edge network.

In~\cite{9040262}, the authors list the following challenges for service migration in federated edges:
\begin{enumerate}[label={\roman*)}]
    \item migration of service functions within and across network domains, such as algorithms to determine when and where to migrate as well as schemes for migration to guarantee that services will not be interrupted;
    \item network update for service function migration, such as, for example, network reconfiguration for migration within a domain and overlay network coordination for inter-domain service chain;
    \item efficient state migration to ensure consistency and inconsistency detection and recovery;
    \item fault-tolerance mechanisms to the reliability of  federated services include  passive or active replication of service function instances, federation of controllers, and a topology-independent loop-free alternate path to  inter-domain and intra-domain service chains.
\end{enumerate}

Facilitating rapid migration between federated edge resources is one of the key challenges for the management of resources. 
To facilitate fast migration, alternate virtualization technologies may need to be developed, which would allow the migration of more abstract entities, such as functions or programs.
Such technology may also find a place in upcoming serverless computing platforms for developing interoperable platforms across federated edge resources.

\section{Conclusion}\label{conclusion}

This paper provides a comprehensive discussion of resource management for edge computing to federated learning, which is a promising solution for reducing latency, and providing privacy of user data as well as for bringing computation closer to the user than traditional machine learning.
However, it brings numerous challenges and raises issues that need to be addressed for the execution of federated learning to execute at the edge.
Discovery, deployment, load balancing, migration, and energy efficiency have challenges that substantially impact the key decision of "Where a machine learning algorithm will be performed".

In this paper, we have provided an overview of the existing solutions in resource management regarding to discovery, deployment, load balancing, migration and energy efficiency.
Moreover, challenges in edge computing and federated learning are discussed.
Future solutions need to focus on efficient approaches for executing FL at the edge, including solutions reducing communication, energy consumption, and latency. 
One of the main challenges facing resource management is the heterogeneity that can arise in the data, hardware, software, and also in communication technology.

To build efficient management solutions, future research needs to focus on all steps of federated learning; it needs to provide a balance between time, energy, and learning, allowing more applications and services to use federated learning.
In this paper, we did not explore security and privacy issues, but federated learning can suffer attacks on the communication between edge devices and edge/cloud servers.
Moreover, communication resources (e.g., 4G, 5G, and WiFi) were not explored in this survey.
For future works, it would be essential to discuss and include suitable solutions for resource management.


\section*{Acknowledgments}
This work was supported by CAPES, CNPq, and grant 15/24494-8, FAPESP, BRAZIL.

\bibliographystyle{unsrt}  
\bibliography{references}

\end{document}